\documentclass[useAMS,usenatbib]{mn2e}
\usepackage[fleqn]{amsmath}
\usepackage{graphicx}
\usepackage{color}

\title[Magnitude and size evolution of bulgeless galaxies]{Magnitude and size evolution of bulgeless galaxies\thanks{Based on observations obtained with the NASA/ESA {\it Hubble Space Telescope}, which is operated by the Association of Universities for Research in Astronomy, Inc.(AURA) under NASA contract NAS 5-26555}}
\author[Sonali Sachdeva]{Sonali Sachdeva$^{1}$\\
$^{1}$Department of Physics and Astrophysics, University of Delhi, Delhi 110007, India}
\begin{document}

\date{in original form 2012 October 2}

\pagerange{\pageref{firstpage}--\pageref{lastpage}} \pubyear{2012}

\maketitle

\label{firstpage}

\begin{abstract}
We examine the magnitude and size evolution of bulgeless (discs with no-bulge or pseudo-bulge) galaxies up to z$\sim$0.9 in rest-frame {\it B}-band. Their evolution is compared to that of normal-discs (or discs with classical-bulge). The study is done for luminous sources ($M_B$$\leq$-20) in two equal-volume redshift-bins (0.4$\leq$z$<$0.77 and 0.77$\leq$z$<$1.0) and a local range (0.02$\leq$z$<$0.05). The mean surface-brightness, $\overline{\mu}_B$, from $z_{mean}$$=$0.89 to $z_{mean}$$=$0.04, shows a dimming of 0.79 mag arcsec$^{-2}$ for bulgeless galaxies and 1.16 mag arcsec$^{-2}$ for normal-discs. The characteristic magnitude, $M_B^*$, shows an increase of 0.55-mag for bulgeless galaxies and 0.95-mag for normal-discs. Both dimming and faintness observed since z$\sim$0.9 is more pronounced for the normal-discs by $\sim$0.4-mag. The size-distribution is log-normal and both bulgeless and normal-discs show a slight increase in the mean value, $\Delta$$\overline{\log(R_e)}$$\sim$0.11, from $z_{mean}$$=$0.89 to $z_{mean}$$=$0.04. The proportion of bulgeless galaxies in the full disc sample undergoes a considerable decline with decrease in redshift. This along with the larger dimming and faintness seen for normal-discs suggests that some fraction of the bulgeless sources switch to the normal-disc morphology with time. To ascertain the validity of studying morphology in the optical, the properties of the galaxies observed in both rest-frame {\it B} and {\it I}-band are compared. The common sample is more luminous in the {\it I}-band but the sizes are larger in the {\it B}-band for more than 74$\%$ of the sources. The variation in the sersic-index values of the galaxies in the two rest-bands is minor enough to have any affect on the morphological classification.
\end{abstract}


\section{Introduction}

The Cold Dark Matter Model based on structure formation through hierarchical clustering given by \citet{b49} and \citet{b15} is the most acceptable scenario of disc formation. Gas (baryons) falling inside the spinning halo of dark-matter, acquires the same distribution of specific angular-momentum as the halo and conserves it while it cools and collapses which leads to the formation of disc inside the halo. 

However, as the haloes grow hierarchically, mergers give rise to dynamical friction which leads to the loss of angular-momentum in these baryons (or formed stars). Additionally, the continuous accretion of intergalactic debris displaces stars from their disc orbits. These stars, which lose their angular momentum due to mergers and accretion, then sink to the centre of galaxies which leads to the formation of centrally concentrated stellar distribution supported by random motion \citep{b21,b10}.   

This centrally concentrated stellar distribution, with an amorphous smooth appearance, is defined as the bulge of the disc galaxy \citep{b54,b11}. The bulge's size compared with the size of the disc is one of the main classification criteria along the spiral sequence from early Sa to Sd, Sm and Im types \citep{b41}. Bulges are, thus, a natural outcome of disc formation in the $\Lambda$CDM scenario \citep{b35}. This is well reflected by the fact that no simulation based on the $\Lambda$CDM paradigm has been able to produce a single galaxy without bulge as yet \citep{b14,b42}.

However, the presence of a large number of giant bulgeless galaxies in the universe challenges this picture of galaxy formation \citep{b26,b27}. The fragile thin disc of stars surviving mergers and accretion without any evidence of merger-built-bulges is potentially catastrophic for the Lambda CDM model \citep{b3,b21,b4}.

The situation gets acute when we realize that many small bulges which were thought to be merger remnants, are found to be pseudo bulges. They are called pseudo-bulges as they have a high ratio of ordered-motion to random-motion and are understood to form mainly due to the secular evolution of isolated galaxy discs \citep{b26}. In terms of cosmology, discs with pseudo-bulges are considered as pure-disc or bulgeless \citep{b27}. The term "bulgeless", thus, refers to disc galaxies which do not have a bulge or which support a pseudo-bulge.

Understanding the formation and evolution of bulgeless galaxies is one of the most challenging problems facing us. To that end, in this paper, we have examined the magnitude and size evolution of bulgeless galaxies in the past $\sim$8 Gyrs. Their evolution is compared to that of normal-disc galaxies. "Normal-disc" galaxies refer to disc galaxies with classical bulge, these discs are easily formed by models based on standard cosmology.

The study is done in Chandra Deep Field-South for the rest-frame {\it B}-band (0.37-0.49 $\mu$m) in two equal volume redshift bins (0.4$\leq$z$<$0.77 and 0.77$\leq$z$<$1.0). The images from the Great Observatories Origins Deep Survey obtained using Advanced Camera for Surveys on the {\it Hubble Space Telescope (HST)} \citep{b20} are combined with the spectrophotometric redshifts from COMBO-17 \citep{b53}.

The study of the distribution of galaxies with respect to luminosity, size, morphological-type and the correlation among them is crucial to our understanding of the formation and evolution of galaxy population \citep{b45}. The correlation for discs has previously been explored \citep{b40,b30,b46,b37,b2,b32,b24}, but the studies have come out with conflicting results.

\citet{b39} found that the mean half-light-radius of spiral galaxies has increased with time by $\sim$25$\%$ since z$\sim$1.2. \citet{b30} found that the size function of large discs is approximately constant to z$\sim$1 and discs are 0.8 mag brighter at z=0.7. Adding to that, \citet{b40} found that the mean rest-frame blue-band surface-brightness for all morphological types increases by 0.95 mag up to $z_{mean}$=0.9. However, \citet{b46} concluded that no discernible evolution remains in the surface-brightness of disc-dominated galaxies once a limit of $M_B$$<$-19 is put on the sample. They emphasized that the evolution seen by previous studies is due to selection-effects in which low-luminosity galaxies at lower redshifts are compared with high-luminosity ones at higher redshifts.

This study was supported by \citet{b37}, who found that for the high central surface-brightness limit $\mu_B^o$$<$20.6 mag arcsec$^{-2}$, no significant evolution is seen in the mean {\it B}-band surface-brightness. Both these studies found a handful of luminous, high surface-brightness galaxies in the highest redshift-bin with a wide range of color and bulge-fractions.

Later, \citet{b2} found a brightening of $\sim$1 mag arcsec$^{-2}$ for disc galaxies in the rest-frame {\it V}-band by z$\sim$1 for $M_V$$\leq$-20. They observed that the previous attempts, with high surface brightness limits, were selecting only those local galaxies which would, in principle, be observed at high redshift. This amounts to studying the very bright tail end of galaxy distribution at all redshifts, and hence, evolution was not seen. This was supported by \citet{b32}. They found a substantial dimming of luminosity, $\Delta M_B$$\sim$1.5-mag, since z=1 for large and intermediate-sized galaxies. Also, \citet{b24} observed that for the size-function of disc galaxies to remain constant since redshift 1.0, the disc galaxies need to go fainter by 1-1.5 mag towards decreasing redshifts.   

In this paper, the disc sample has been divided into two groups, bulgeless and normal-discs. Kormendy-relation along with the sersic-index criteria is used to make the division. The magnitude, size and surface-brightness evolution of the two morphological types is examined separately. The major aim is to see if the bulgeless galaxies have evolved any way differently from their counterparts with bulge. The entire disc sample (0.4$\leq$z$<$1.0) has $M_B$$\leq$-20, taking care of the accuracy limit of redshifts from COMBO-17 and the imaging efficiency of the {\it HST} at the high redshift end. The NYU-Value Added Galaxy Catalog \citep{b8} is used to obtain a local sample (0.02$\leq$z$<$0.05) of bulgeless and normal-disc galaxies in the rest-frame {\it B}-band. This is viewed as a reference to the situation at the present epoch.  
   
\citet{b14} argued that the surface-brightness profile in optical light may not reliably trace the surface-density profile in stellar mass. Also, {\it I}-band should study the surface-density profile better as it will be relatively free of biases of young stellar population and dust. To explore the differences in surface-brightness and surface-density profile, the luminosity, size and morphology of the galaxies observed in rest-frame {\it B} and {\it I}-band are compared. The rest-frame {\it I}-band (0.78-0.85 $\mu$m) properties of the galaxies (for 0.4$\leq$z$<$1.0) are obtained from the images from {\it HST} Wide Field Camera-3 (WFC3) Early Release Science (ERS) programme in the CDF-S in the near-IR filters \citep{b52}. 

We consider a flat $\Lambda$-dominated universe with $\Omega_{\Lambda}$=0.73, $\Omega_m$=0.27, H$_o$=71 km sec$^{-1}$ Mpc$^{-1}$. In Section 2, the criteria for selecting the sample and identifying the morphology are described. In Section 3, surface-brightness evolution of the sample is examined. In Section 4, luminosity evolution is studied via luminosity-function and size evolution is studied via log-normal-curves. In Section 5, properties of the galaxies are compared in rest-frame {\it B} and {\it I}-band. A discussion of the results and conclusions of the study are presented in Section 6. 

\section{Sample selection and identifying the morphology}

\subsection{Sample selection}

\subsubsection{Obtaining catalogues}

The images in {\it V} (F606W), {\it i} (F775W) and {\it z} (F850LP) filters of ACS ({\it HST}) are used to study GOODS CDF-S in rest-frame {\it B}-band. The SExtractor software \citep{b5} is used for source detection. The source detection is based on the F850LP image. The postage stamp size of the objects, to obtain cut-outs, is determined by using the SExtractor output parameters, mainly KRON-RADIUS. The sky background for each source is estimated by using the flux growth curve method on the whole image masking out the neighbouring sources. Galfit \citep{b36}, which performs the two-dimensional modelling of the surface-brightness distribution of the galaxies, is employed for extracting galaxy parameters. The SExtractor output parameters (MAG-BEST, FLUX-RADIUS, etc) are given as initial parameters to Galfit, which performs a sersic profile fitting on all sources.

The sersic profile (S\'ersic 1968) for the variation of galaxy's surface brightness from it's centre is given as:
\begin{equation} 
I(r)=I_b(0)\exp(-2.303b_n(r/r_e)^{1/n}),
\end{equation}
where n is the sersic-index which controls the degree of curvature of the profile. $I_b(0)$ is bulge central intensity and the constant $b_n$ is chosen such that $r_e$ is half-light-radius for every value of n. 

Galfit uses the Levenberg-Marquardt algorithm to do the sersic profile fitting. We obtain integrated magnitude, half-light-radius, sersic-index and axis-ratio of the galaxies, along with the associated errors. Only the sources with reduced $\chi^2$-value of less than 5 are selected. Redshifts are then obtained from the COMBO-17 survey in CDF-S \citep{b53}. They are accurate to 1$\%$ in $\frac{\delta z}{(1+z)}$ at R$<$21. To obtain the redshifts, RA and Dec were matched for a maximum distance of 0.00014 deg which corresponds to 0.5 arc-seconds. This strikes a balance such that maximum number of sources obtain a redshift and the cases of two neighbouring sources getting the same redshift is reduced to minimal. 

The catalogues obtained in {\it V}, {\it i} and {\it z} filters are averaged and merged according to redshift-ranges to create a rest-frame {\it B}-band catalog. It has 5103 sources in the redshift-range 0.4 to 1.0. The images in near-IR filters, {\it J} (F125W) and {\it H} (F160W) of WFC3 ({\it HST}) in the CDF-S are used in a similar manner to obtain the properties in rest-frame {\it I}-band. The source detection is based on the F160W image. After obtaining the cut-outs and estimating the sky-background, sersic component fitting is done on the sources. The final catalog in the redshift-range 0.4 to 1.0 has 1109 sources. It is used in Section~5 to compare the morphology, luminosity and sizes of the galaxies in the two rest-frame bands.

Some entries for the {\it B}-band catalog are shown in Table~1. Integrated-magnitude (m), half-light-radius (in pixels) ($r_e$), sersic-index (n), axis-ratio (ar) and the errors associated are as found from the sersic-component fitting. The redshift (z) and the error associated is from the COMBO-17 survey columns (MC-z and e-MC-z). Both {\it B} and {\it I}-band catalogues are made available with the on-line version.

\begin{table*}
 \centering
 \begin{minipage}{200mm}
  \caption{The {\it B}-band catalog in the 0.4-1.0 redshift-range (for few sources)}
  \begin{tabular}{@{}lllllllllllll@{}}
  \hline
   ID & m & m-err & $r_e$ & $r_e$-err & n & n-err & ar & ar-err & RA & Dec & z & z-err\\
 \hline
 10135 & 23.1481 & 0.0032 & 12.6368 & 0.0673 & 0.4707 & 0.0053 & 0.859 & 0.0037 & 53.0851141 & -27.7289909 & 0.521641 & 0.0364272\\   
 10154 & 24.9811 & 0.0104 & 10.8213 & 0.1491 & 0.5772 & 0.0312 & 0.1489 & 0.0035 & 53.0852279 & -27.7538798 & 0.597967 & 0.178344\\    
 10150 & 25.2588 & 0.0322 & 4.9653 & 0.2884 & 2.9811 & 0.2465 & 0.3941 & 0.0178 & 53.085196 & -27.826633 & 0.453996 & 0.368795\\    
 10170 & 23.0068 & 0.0158 & 27.1952 & 0.7499 & 3.5955 & 0.0642 & 0.2823 & 0.0026 & 53.0853214 & -27.7923126 & 0.581040 & 0.0443867\\   
 10241 & 23.9253 & 0.0061 & 9.6979 & 0.0914 & 1.1951 & 0.018 & 0.2245 & 0.0019 & 53.0857339 & -27.895439 & 0.478985 & 0.0578052\\   
 10263 & 24.3276 & 0.0156 & 11.5209 & 0.2698 & 1.3609 & 0.0416 & 0.2807 & 0.0047 & 53.0858755 & -27.7059257 & 0.498965 & 0.0333549\\   
 10290 & 26.4758 & 0.0483 & 7.6923 & 0.5101 & 1.0127 & 0.1378 & 0.2892 & 0.0161 & 53.0860538 & -27.7855397 & 0.526688 & 0.368991\\    
 10486 & 22.4624 & 0.0029 & 18.0997 & 0.0751 & 0.7076 & 0.0055 & 0.3485 & 0.001 & 53.0873424 & -27.7430061 & 0.520664 & 0.0321240\\   
 10624 & 23.9292 & 0.0054 & 10.5075 & 0.085 & 0.7559 & 0.014 & 0.3075 & 0.002 & 53.0881005 & -27.7530047 & 0.515495 & 0.0351864\\   
 10707 & 20.9032 & 0.0146 & 63.1796 & 1.7937 & 7.8752 & 0.069 & 0.7586 & 0.0024 & 53.0886768 & -27.7432276 & 0.451917 & 0.0412592\\      
\hline
\end{tabular}
\end{minipage}
\end{table*} 

\subsubsection{Applying limits}

In the {\it B}-band catalog, the sources with more than 30$\%$ error in radius ($\frac{\delta r}{r}$$>$0.3) and sersic-index ($\frac{\delta n}{n}$$>$0.3) are removed. Also, to remove spurious sources, the conditions 0.3$<$$r_e$$<$300 (pixels) and 0.1$<$n$<$8.0 are applied. This reduces the number of sources in the rest-frame {\it B}-band to 4124 for the redshift range 0.4-1.0. The integrated-apparent-magnitude and half-light-radius (pixels) are converted into absolute-magnitude and half-light-radius (Kpc) as per the redshift and cosmology considered. The relations used for these conversions are explained in Section~3. The absolute-magnitudes obtained are comparable with the absolute-magnitudes obtainable from the COMBO-17 survey \citep{b53} for the rest-frame {\it B}-band.

The redshift-magnitude distribution in 0.4-1.0 z-range for galaxies with $M_B$$>$-20 and $M_B$$\leq$-20 is shown in Fig.~1. The plot with $M_B$$>$-20 shows that galaxies with lower luminosity are not seen at high redshift ranges at all. The dashed line made at z$=$0.77 separates the full redshift range (0.4-1.0) into two equal comoving-volume bins. The plot with $M_B$$\leq$-20 shows that the number of galaxies in the two bins is almost same. The magnitude limit for reliable redshift from COMBO-17 is $m_Z$$\sim$23.5. For our upper redshift limit of z$=$1.0, this corresponds to $M_B$$\sim$-20. Based on the depth of {\it HST} imaging, and the redshift accuracy limit of COMBO-17, a magnitude cut of -20 is applied on the sample. We obtain 727 sources in the rest-frame {\it B}-band (0.4$\leq$z$<$1.0) with $M_B$$\leq$-20.   

\subsection{Identifying the morphology}

\subsubsection{Sersic-index}

The sersic-index (S\'ersic, 1968) is considered an elegant tool to determine morphology. The smaller the value of n, the less centrally concentrated the profile and shallower the logarithmic slope at small radii \citep{b36}. The sersic-index value n$=$2.5 is considered an efficient separator of early-type (n$>$2.5) and late-type (n$<$2.5) galaxies, through simulations and visual verifications \citep{b2,b48}. We have, thus, used the criteria to obtain 496 late-type or disc dominated (n$<$2.5) galaxies in the rest-frame {\it B}-band (0.4$\leq$z$<$1.0) with $M_B$$\leq$-20. 

To obtain a sample of bulgeless galaxies, we need to separate discs with no-bulge or pseudo-bulge from our disc dominated sample. Previous studies \citep{b29,b16,b45} have found the sersic-index limit values ranging from 1.5 to 2.2 to separate the bulgeless sample. 

To determine the sersic-index limit, the entire sample of the rest-frame {\it B}-band (i.e. without the magnitude cut, 4124 sources) was divided into three sersic-index ranges (0.8$>$$n_B$, 0.8$\leq$$n_B$$<$1.7, 1.7$\leq$$n_B$) with each range getting almost equal number of sources. For each index range, 0.5-mag size absolute-magnitude bins were created and mean half-light-radius ($R_{eB,m}$) was found for each bin. The distribution of these means against the absolute-magnitude bins for different sersic-index ranges is shown on a log scale in Fig.~2.

The findings support those of \citet{b45}, such that the sersic-index value of 1.7 indeed seems to separate the galaxies into two groups which follow different $\overline{R}_{eB}$-$M_B$ relations, independent of n. The cut, thus, appears to separate galaxies which follow an exponential surface-brightness profile (Sb/Sc) from those which do not. The second plot (Fig.~2) however shows that the scatter is quite large.

\subsubsection{Kormendy-relation}

To ascertain the separation of bulgeless galaxies, Kormendy-relation is applied in addition to the sersic-index criteria. \citet{b19} suggested that the Kormendy-relation which is followed by elliptical galaxies can be used to separate pseudo-bulge (or bulgeless) galaxies. The pseudo-bulge galaxies will show themselves as outliers to the relation.

Galaxies with 2.8$\leq$$n_B$$<$4.5 from the 727 sources (0.4$\leq$z$<$1.0, $M_B$$\leq$-20) were selected as ellipticals \citep{b31}. The equation obtained from the linear-fit to the surface-brightness and log-size data of these ellipticals is the Kormendy-relation. The relation is obtained separately for the two equal volume redshift-bins (0.4$\leq$z$<$0.77 and 0.77$\leq$z$<$1.0), to account for evolution, if any. They are shown as solid lines in the two plots of Fig.~3. The two dashed lines are as obtained from the $\pm$3$\sigma$ values of the zero-point with fixed slope. Only disc dominated ($n_B$$<$2.5) sources have been shown. The discs with pseudo-bulge are outliers in the two plots (found below the lower line) i.e. they follow the relations: 
\begin{equation}
\mu_{e,B}>19.36+2.92*\log(R_{e,B}),\; \; \forall \: 0.4\leq z<0.77
\end{equation}
\begin{equation}
\mu_{e,B}>19.32+2.92*\log(R_{e,B}),\; \; \forall \: 0.77\leq z<1.0
\end{equation}
            
\subsubsection{Division and bulge-total ratio}

More than 85$\%$ of the galaxies in the lower bin (0.4-0.77) and more than 80$\%$ of the galaxies in the higher bin (0.77-1.0)  which are found to have pseudo-bulges (or no bulges) according to these relations, have $n_B$$<$1.7. The two criteria are thus complementary to each other. Only those galaxies which satisfy both criteria, i.e. have sersic-index less than 1.7 and are outliers to the Kormendy-relation, are chosen to be bulgeless (i.e. these discs either have no-bulge or have a pseudo-bulge). They are marked as solid squares in Fig.~3.

The rest of the discs, which either do not follow the relations mentioned above or have 1.7$\leq$$n_B$$<$2.5, are normal-discs i.e. disc galaxies with classical-bulge component. They are marked as open circles in Fig.~3. This sample is selected alongside for comparison. 

To be able to comment on the bulge-total ratio of the sample, we need to know the Petrosian-concentration ($r_{90}$/$r_{50}$) since it is considered a good predictor of the ratio \citep{b19,b28}. 

\citet{b7} calculated the Petrosian inverse concentration ($r_{50}$/$r_{90}$) dependence on the sersic-index, for different values of axis-ratios. Our bulgeless sample has sersic-index, n, less than 1.7 and axis-ratio, b/a, is less than 0.6 for more than 90$\%$ of the sample. Thus according to \citet{b7} (see their Fig.~5), the Petrosian inverse concentration ($r_{50}$/$r_{90}$) is more than 0.36 for our bulgeless sample. 

The bulge-total ratio scales linearly with the Petrosian concentration. For $r_{50}$/$r_{90}$$>$0.36 and b/a$<$0.6, the bulge-total ratio is found to be less than 0.3 (or 30$\%$) for our bulgeless sample (Fig.~5 of \citet{b7}). The bulge-total ratio is evidently less than 0.2 for most of the bulgeless sources as the average values of sersic-index and axis-ratio of the sample are both quite lower than their upper limits of 1.7 and 0.6 respectively.  

Out of the 496 disc dominated galaxies with $M_B$$\leq$-20 and 0.4$\leq$z$<$1.0 in the rest-frame {\it B}-band, 186 are found to be bulgeless and 310 are normal-discs. 

\subsection{Local sample}

The New York University-Value Added Galaxy Catalog (NYU-VAGC) maintained for the study of galaxy formation and evolution \citep{b8} is used to obtain a local sample for reference. The NYU-VAGC is a catalog of local galaxies based on a set of publicly released surveys matched to the Sloan Digital Sky Survey (SDSS) Data Release 2 \citep{b1}. It has a specially prepared catalog for low-redshift galaxies (0.0033$<$z$<$0.05) \citep{b9}.

We use the {\it g} and {\it r}-band data to create a rest-frame {\it B}-band catalog. The magnitudes are galactic-extinction corrected \citep{b44} and K-corrected \citep{b6} to the rest-frame band passes. We use the relations given by \citet{b18} to obtain absolute-magnitudes in rest-frame {\it B}-band. The sersic half-light-radii of the {\it g} and {\it r}-band are converted from arc-seconds to Kpc according to their redshift. The relations given by \citet{b2}, that were found using the \citet{b13} data to account for the radial color gradients, are used to obtain half-light-radii in the rest-frame {\it B}-band. There were a large number of sources with negligible values of the parameters ($r_e$$<$0.1 arc-seconds) in the highly local redshift-range (0.0033$\leq$z$<$0.02) of the catalog. We obtain 43,919 sources in the rest-frame {\it B}-band in the redshift-range 0.02$\leq$z$<$0.05. The magnitude cut of $M_B$$\leq$-20, reduces the number of sources substantially. The final catalog is of 764 galaxies in rest-frame {\it B}-band, in the redshift-range 0.02$\leq$z$<$0.05 with $M_B$$\leq$-20. A sample of bulgeless and normal-discs is then obtained using sersic-index and Kormendy-relation criteria.

We thus have samples of bright ($M_B$$\leq$-20), bulgeless and normal-disc galaxies, in two equal volume redshift bins (0.4$\leq$z$<$0.77 and 0.77$\leq$z$<$1.0), and in a local redshift-range (0.02$\leq$z$<$0.05) for rest-frame {\it B}-band. This adds up to a total of 597 galaxies, of which 211 are bulgeless and 386 are normal-discs. 

\begin{figure}
\mbox{\includegraphics[width=65mm]{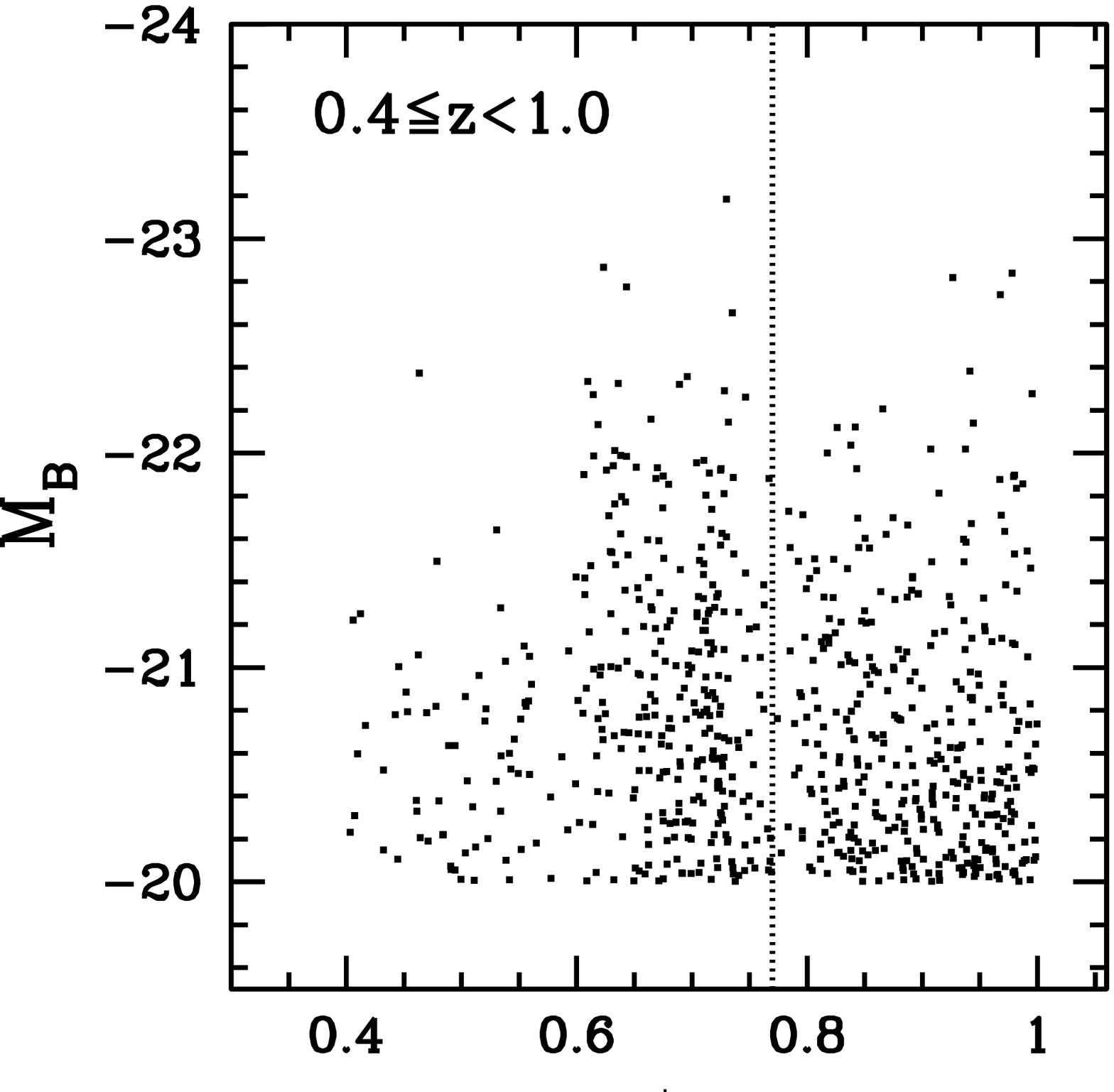}}
\mbox{\includegraphics[width=65mm]{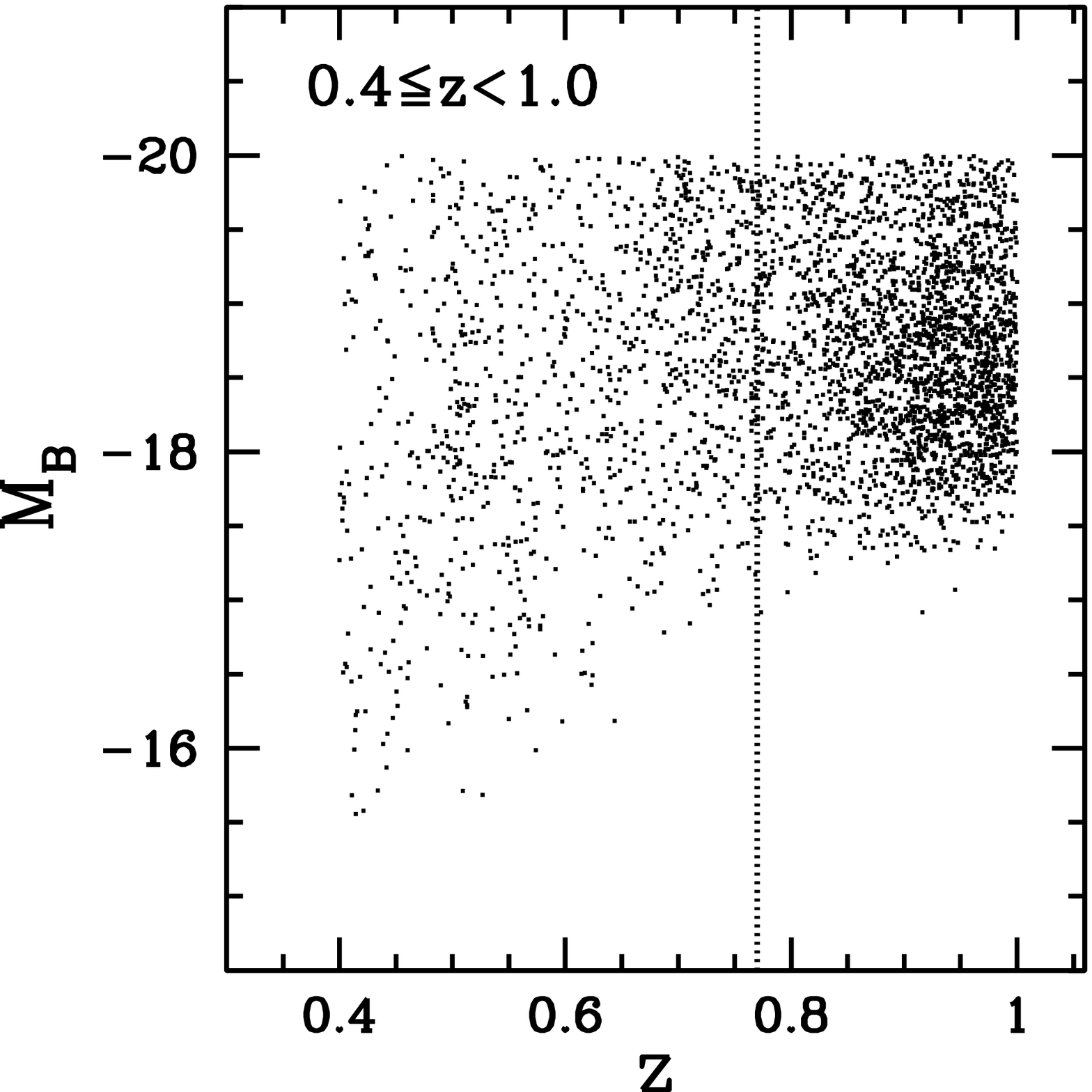}}
  \caption{The redshift-magnitude distribution is shown in 0.4-1.0 z-range for $M_B$$\leq$-20 and $M_B$$>$-20. The distribution for $M_B$$>$-20 shows that the low-luminosity galaxies are not observed at high redshifts. The dashed line at z$=$0.77 separates the redshift range (0.4-1.0) into two equal volume redshift bins. The distribution for $M_B$$\leq$-20 shows that the number of sources are almost equal in the two bins.}
\end{figure}

\begin{figure}
\mbox{\includegraphics[width=70mm]{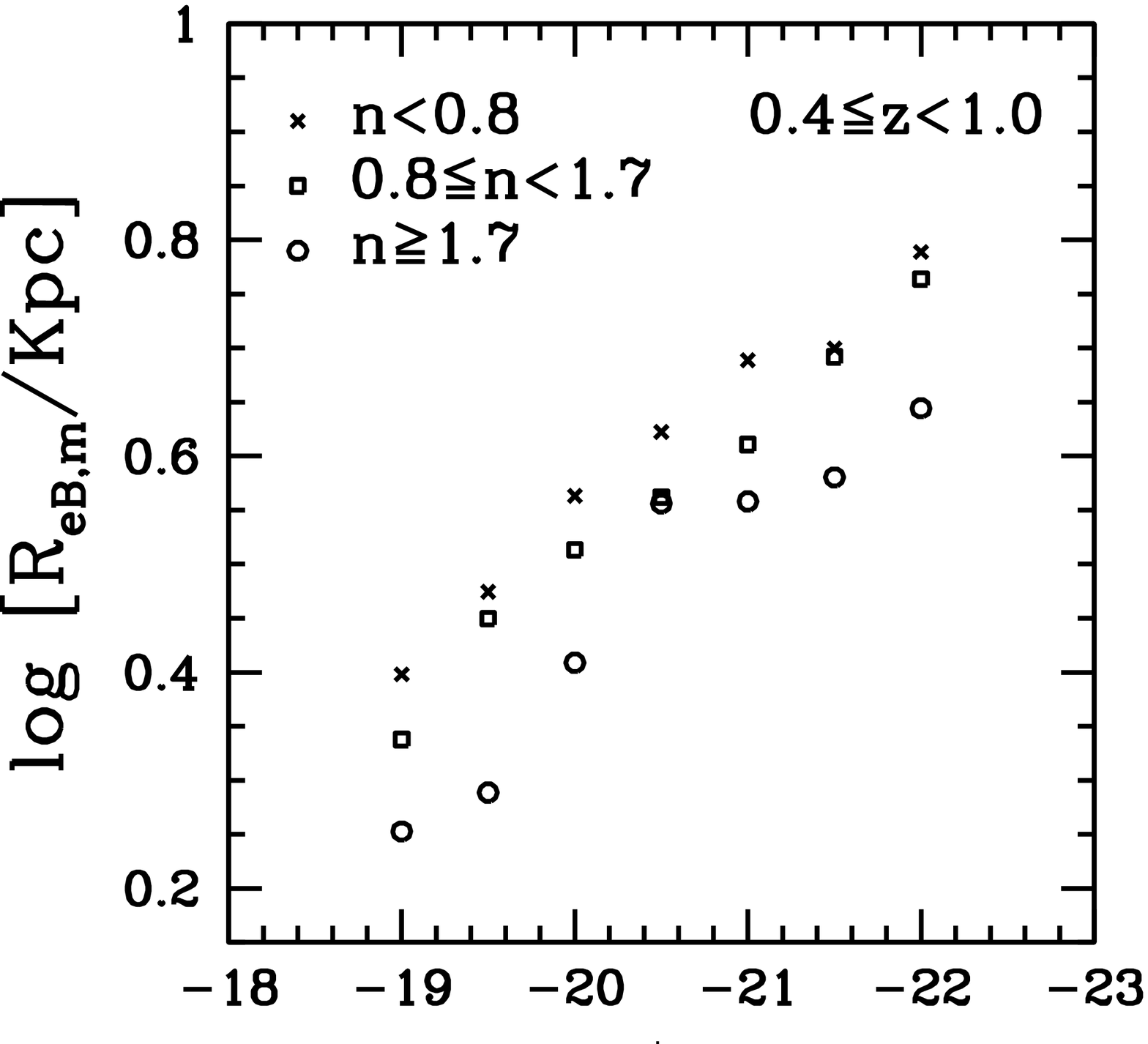}}
\mbox{\includegraphics[width=70mm]{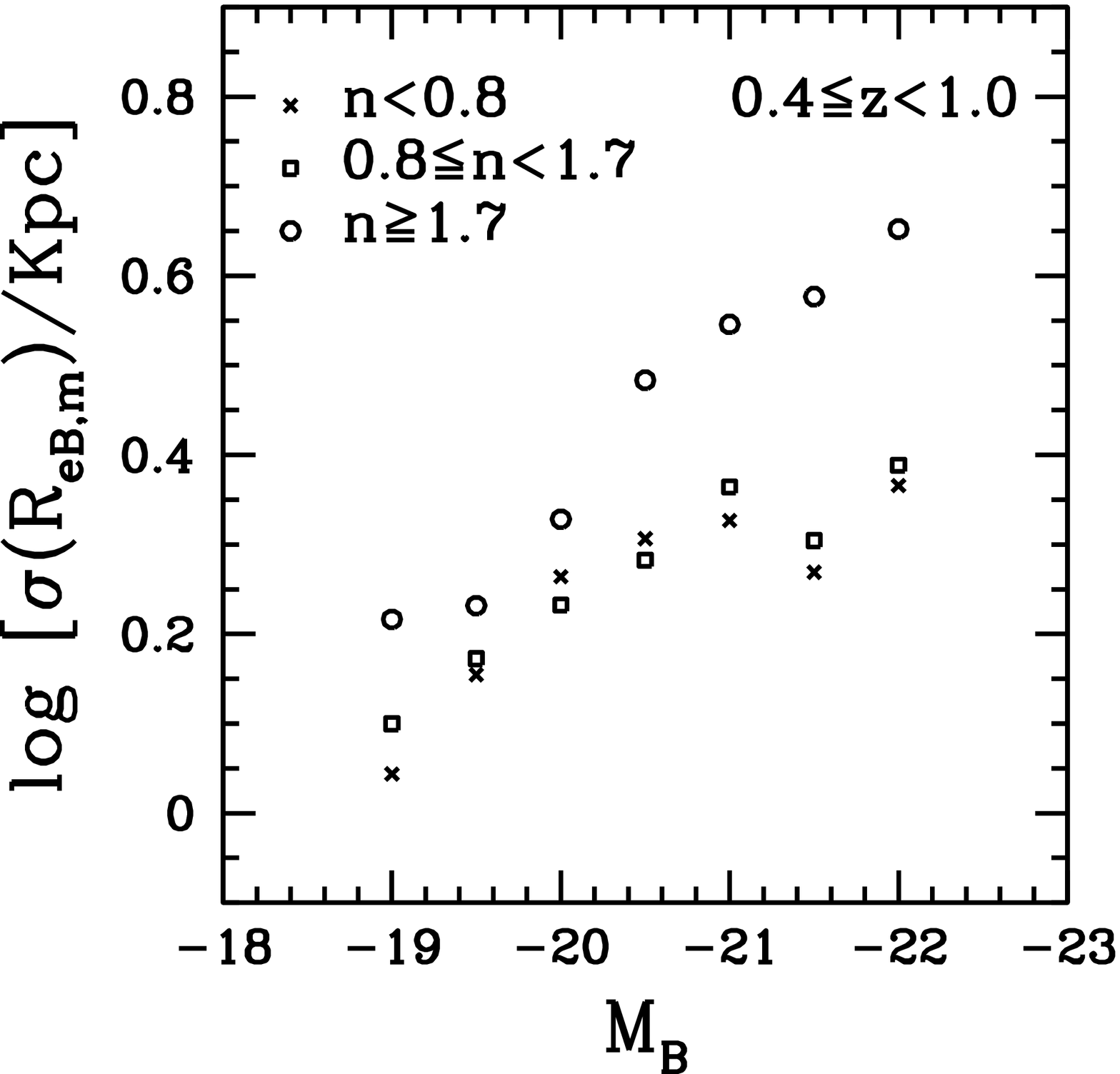}}
  \caption{The distribution of the means of the half-light-radii in different magnitude bins is shown on a log-scale for galaxies belonging to different ranges of sersic-index. The data points for $n_B$$>$1.7 follow a separate relation as they represent galaxies with bulge. The standard-deviations associated with the mean values are also shown for the three sersic ranges which show that the scatter is quite large.}
\end{figure}

\begin{figure}
\mbox{\includegraphics[width=70mm]{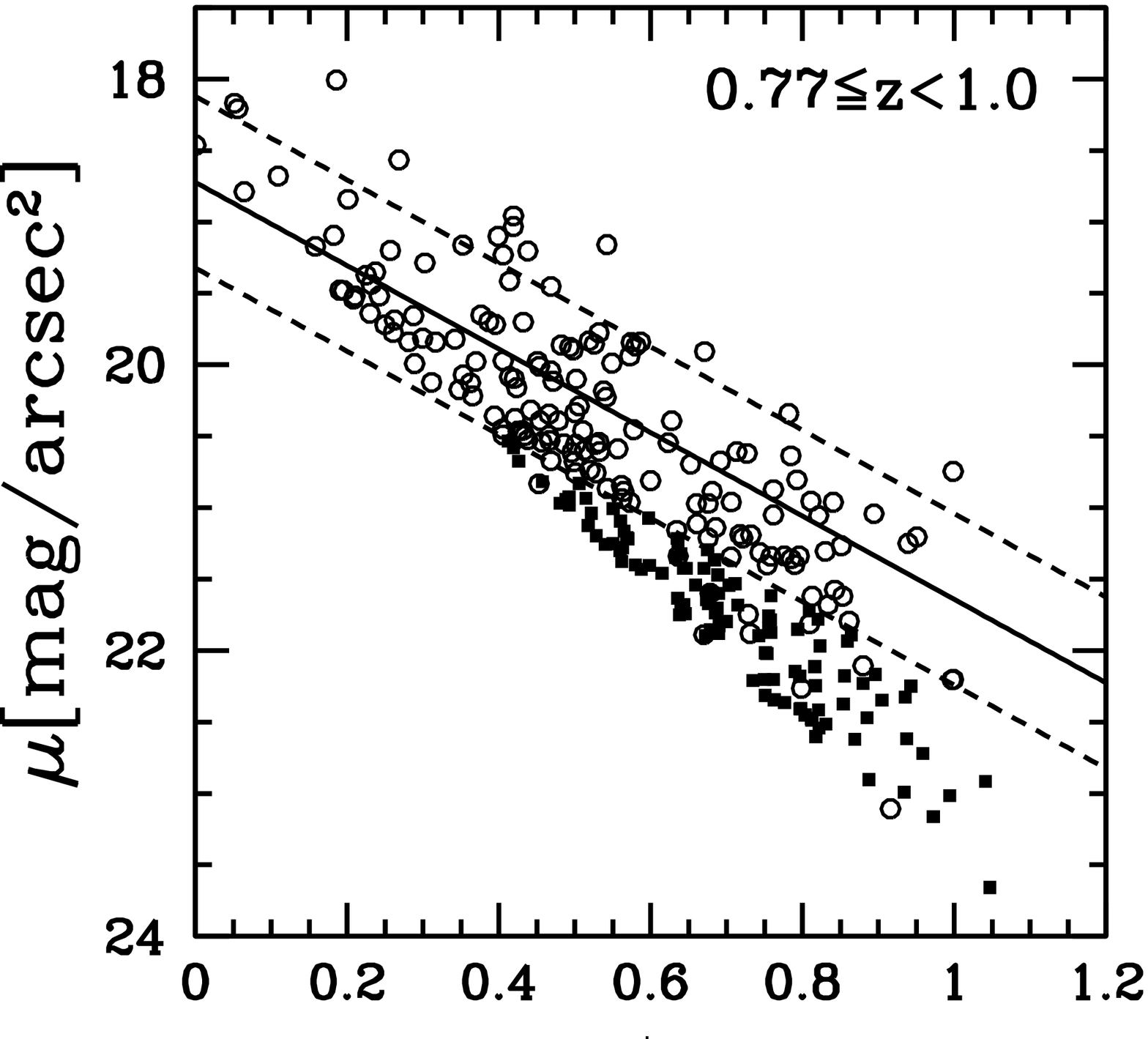}}
\mbox{\includegraphics[width=70mm]{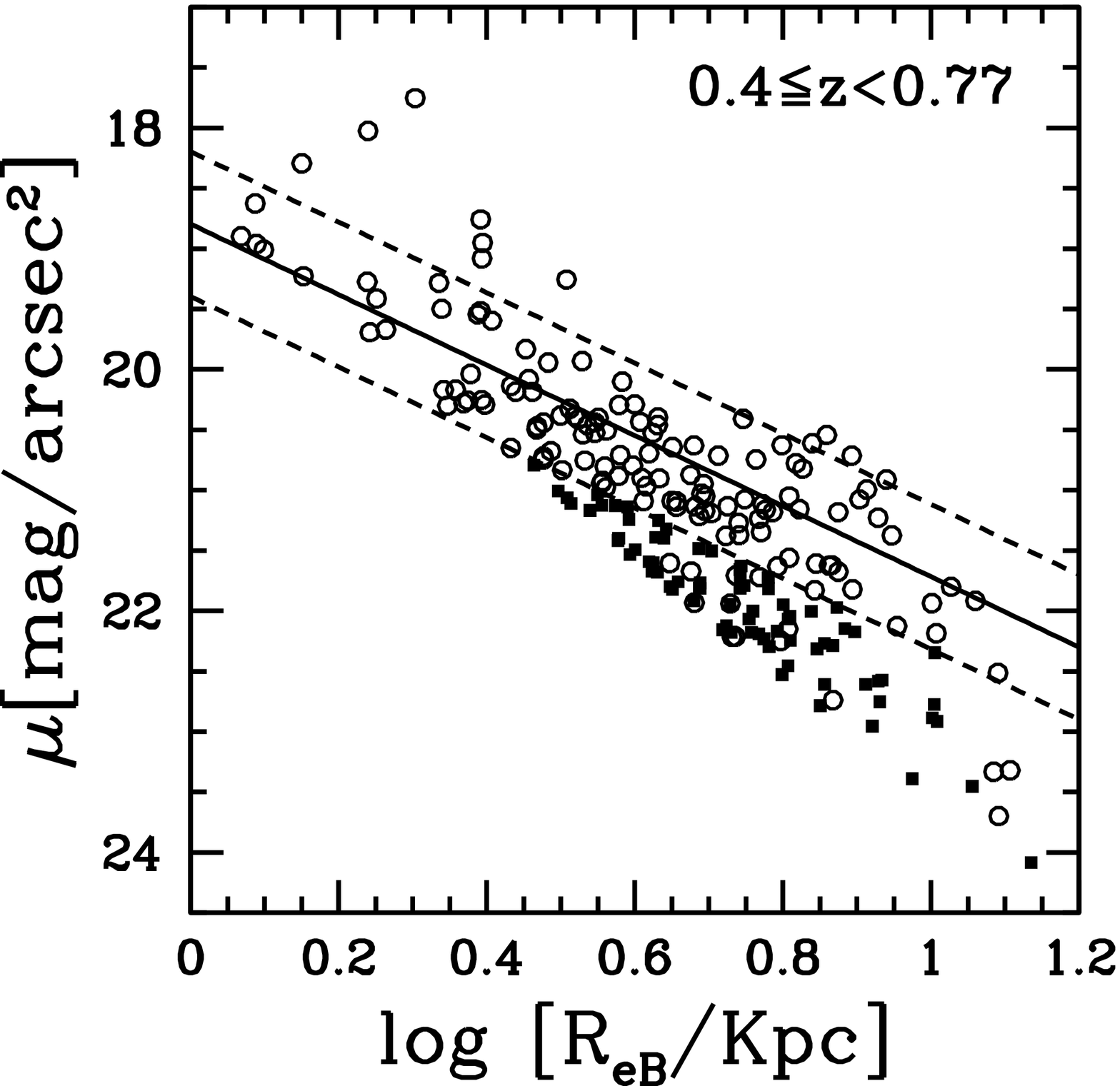}}
  \caption{The surface-brightness and log-size distribution of the disc dominated galaxies is shown in the two redshift ranges. The Kormendy-relation, obtained for the elliptical sources, is shown as the solid-line. The dashed lines mark the $\pm$3$\sigma$ scatter associated with the zero-point value, obtained for a fixed slope. The outliers, i.e. those lying below the lower dashed-line, are selected as bulgeless if they also have $n_B$$<$1.7 (marked as solid squares). The rest of the sources are selected as normal-discs (marked as open circles).}
\end{figure}

\section{Surface-brightness distribution and evolution}

\subsection{Intrinsic values}

The absolute-magnitude of the galaxies in rest-frame {\it B}-band, $M_B$, is calculated from their apparent-magnitude, m, using the relation:
\begin{equation}
M_B=m-5*log_{10}(D_L*10^5)-K,
\end{equation}
where $D_L$ is the luminosity-distance in Mpc, calculated using redshift of the galaxies as per chosen cosmology. K is the K-correction term to account for the fact that the band in which apparent magnitude is measured is different from the rest-frame band. The sign convention is according to \citet{b33}. The K-correction depends on the object's SED \citep{b22}. For a power-law continuum, it is given by the relation:
\begin{equation}
K_{cont}=-2.5*(1+\alpha_{\nu})*log_{10}(1+z),
\end{equation}
where $\alpha_{\nu}$ is the slope of the continuum with canonical value of -0.5 \citep{b38}. Thus the overall equation becomes:
\begin{equation}
M_B=m-5*log_{10}(D_L*10^5)+2.5*log_{10}(\sqrt(1+z))
\end{equation}

From the sersic-catalogues, we have the half-light-radius of the galaxies in pixels. The pixels are converted into arc-seconds according to the plate-scale of the telescope, and are then converted into radians. The intrinsic half-light radius (or effective-radius) in rest-frame {\it B}-band, $R_{eB}$, within which half of the total intensity of the galaxy is contained (see Eqn.~1), is then computed (in Kpc) using:
\begin{equation}
R_{eB}=D_A*1000*\Delta\Theta,
\end{equation}
where $D_A$ is the angular-diameter-distance in Mpc calculated using redshift of the galaxies as per chosen cosmology and $\Delta\Theta$ is the radians covered on detector.

The intrinsic surface-brightness, in mag arcsec$^{-2}$, is thus calculated using the relation:
\begin{equation}
\mu_B=M_B+5*log_{10}R_{eB}+38.568,
\end{equation}
where $M_B$ is in mag and the constant term accounts for the fact that $R_{eB}$ is in Kpc. 

\subsection{Magnitude-size distribution}

The absolute-magnitude and half-light-radius distribution of the bulgeless and normal-disc galaxies in the three redshift bins (0.77$\leq$z$<$1.0, 0.4$\leq$z$<$0.77 and 0.02$\leq$z$<$0.05) is shown in Fig.~4. The solid line is the Freeman's relation \citep{b17} drawn using the constant surface-brightness value (21.65 mag arcsec$^{-2}$). For the disc-sample as a whole, the scatter is almost same on both sides of the solid line. However, the bulgeless sources populate the upper part and the normal-discs populate the lower part of the constant surface-brightness line. The bulgeless seem to support larger sizes as compared to the normal-discs and there is a deficit of bulgeless sources at low magnitudes (high luminosities).

\subsection{Surface-brightness distribution}

The surface-brightness distribution of bulgeless and normal-disc galaxies in the three redshift bins (0.77$\leq$z$<$1.0, 0.4$\leq$z$<$0.77 and 0.02$\leq$z$<$0.05) is shown in Fig.~5. Lines passing through the mean surface-brightness ($\overline{\mu}_B$) values of bulgeless and normal-discs sources in the higher z-range plot, are drawn as it is in the lower ranges plots. The solid-dot indicates the mean of bulgeless sources, in each redshift-range, while the open-circle indicates the mean of normal-discs. A shift towards higher $\overline{\mu}_B$ (or lower brightness) can be seen for both the morphological types. In the local (0.02-0.05) plot, it is evident that the shift is more for the normal-discs than for the bulgeless. 

\subsection{Surface-brightness evolution}

The mean values of the surface-brightness of bulgeless and normal-discs for the three redshift-ranges are summarized in Table~2. For the bulgeless galaxies, the dimming is of 0.17 mag arcsec$^{-2}$ from higher (0.77-1.0) to lower (0.4-0.77) redshift-range. From there to the local (0.02-0.05) redshift-range, there is a 0.63 mag arcsec$^{-2}$ dimming. The overall evolution from $z_{mean}$$=$0.89 to $z_{mean}$$=$0.04, for bulgeless galaxies in rest-frame {\it B}-band is of 0.79 mag arcsec$^{-2}$.

For the normal-discs, the dimming from higher (0.77-1.0) to lower (0.4-0.77) redshift-range is of 0.35 mag arcsec$^{-2}$. From lower (0.4-0.77) to the local (0.02-0.05) redshift-range, there is a 0.81 mag arcsec$^{-2}$ dimming. The overall evolution from $z_{mean}$$=$0.89 to $z_{mean}$$=$0.04, for normal-discs in rest-frame {\it B}-band is of 1.16 mag arcsec$^{-2}$. 

The surface-brightness dimming seen for normal-discs (1.16 mag arcsec$^{-2}$) is more than that seen for the bulgeless galaxies (0.79 mag arcsec$^{-2}$) by $\sim$0.4 mag. 

For the disc sample as a whole (Table~2), the overall dimming from $z_{mean}$$=$0.89 to $z_{mean}$$=$0.04 is of 0.86 mag arcsec$^{-2}$. The above deductions show that the contribution to the dimming is more from the normal-discs than from the bulgeless sources. 

\subsection{Space-density and proportion}

The number and comoving-volume of the two morphological types is also mentioned in the table for different z-ranges. The comoving volume is found according to the redshift-range probed and cosmology used. Bulgeless sources in the local redshift-range have a very small space-density as compared to the other two redshift-ranges. Also, the proportion of bulgeless galaxies in the full disc sample declines with decreasing redshift. For high-z bin (0.77-1.0), 39.7$\%$ of the discs are bulgeless, for low-z bin (0.4-0.77), it reduces to 34.7$\%$ and for local-z bin (0.02-0.05), it further reduces to 24.8$\%$.

The change in morphology for some fraction of the bulgeless galaxies, as we approach the present epoch, may explain the considerable decline in their space-density and the decrease in their proportion with respect to the full disc sample. It can also be the probable reason for the larger amount of dimming seen in normal-discs as compared to the bulgeless galaxies.      

\begin{table}
 \centering
 \begin{minipage}{100mm}
  \caption{The mean surface-brightness in rest-frame {\it B}-band}
  \begin{tabular}{@{}llllllllll@{}}
  \hline
   Redshift & Disc & No. of & Comoving & Surface-brightness\\
   range & type & sources & volume & $\overline{\mu}_B$, $\sigma(\overline{\mu}_B)$\\
   z & & & Mpc$^3$ & mag arcsec$^{-2}$\\
 \hline
 0.77-1.0 & bulgeless & 110 & 7.529e+4 & 21.788, 0.599\\
 0.4-0.77 & bulgeless & 76 & 7.307e+4 & 21.959, 0.614\\
 0.02-0.05 & bulgeless & 25 & 1.9256e+6 & 22.586, 0.627\\
 0.77-1.0 & normal & 167 & 7.529e+4 & 20.359, 0.893\\
 0.4-0.77 & normal & 143 & 7.307e+4 & 20.713, 1.011\\
 0.02-0.05 & normal & 76 & 1.9256e+6 & 21.518, 0.573\\
 0.77-1.0 & full-disc & 277 & 7.529e+4 & 20.929, 1.055\\
 0.4-0.77 & full-disc & 219 & 7.307e+4 & 21.147, 1.072\\
 0.02-0.05 & full-disc & 101 & 1.9256e+6 & 21.785, 0.746\\
\hline
\end{tabular}
\end{minipage}

\end{table}

\begin{figure*}
\mbox{\includegraphics[width=53.5mm]{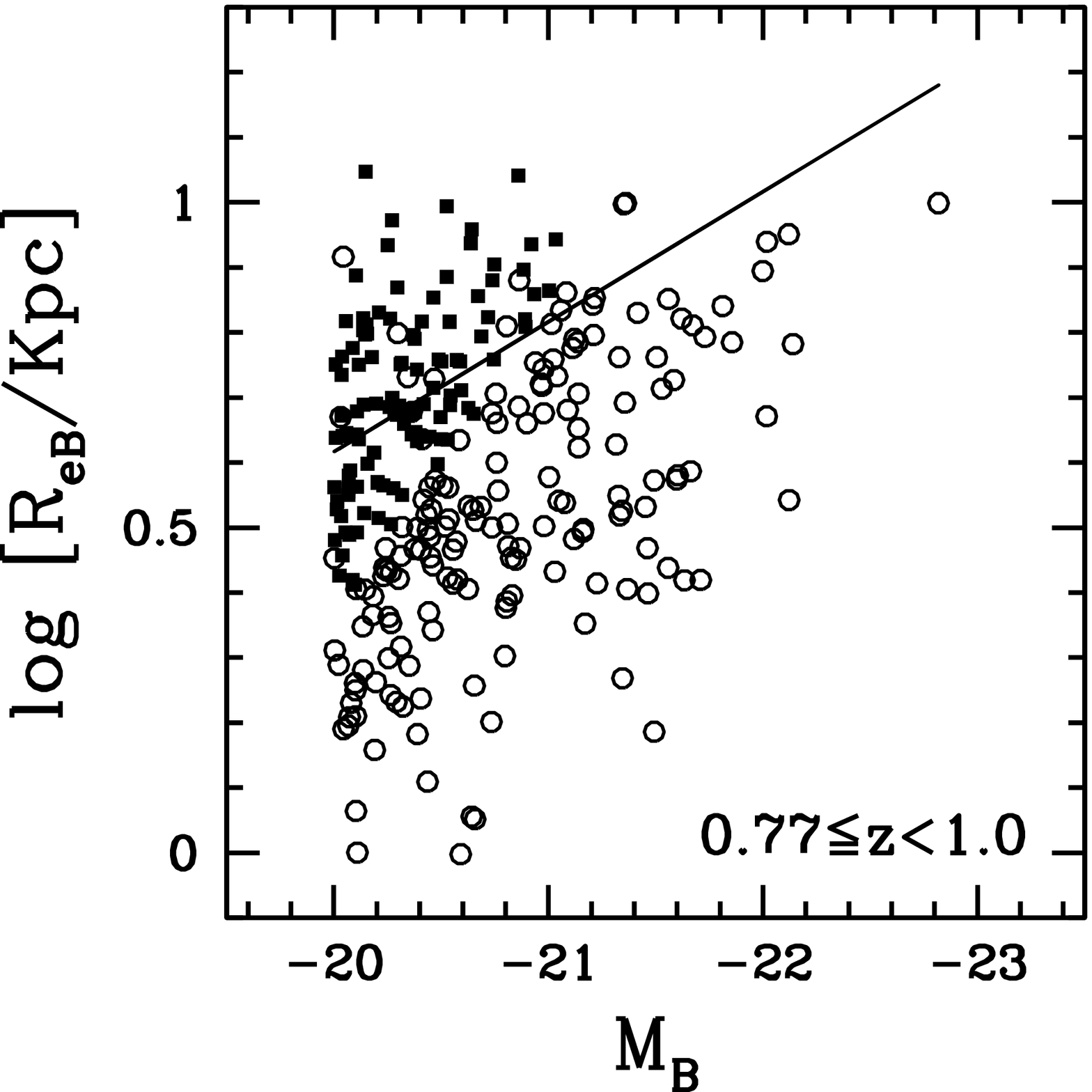}}
\mbox{\includegraphics[width=50mm]{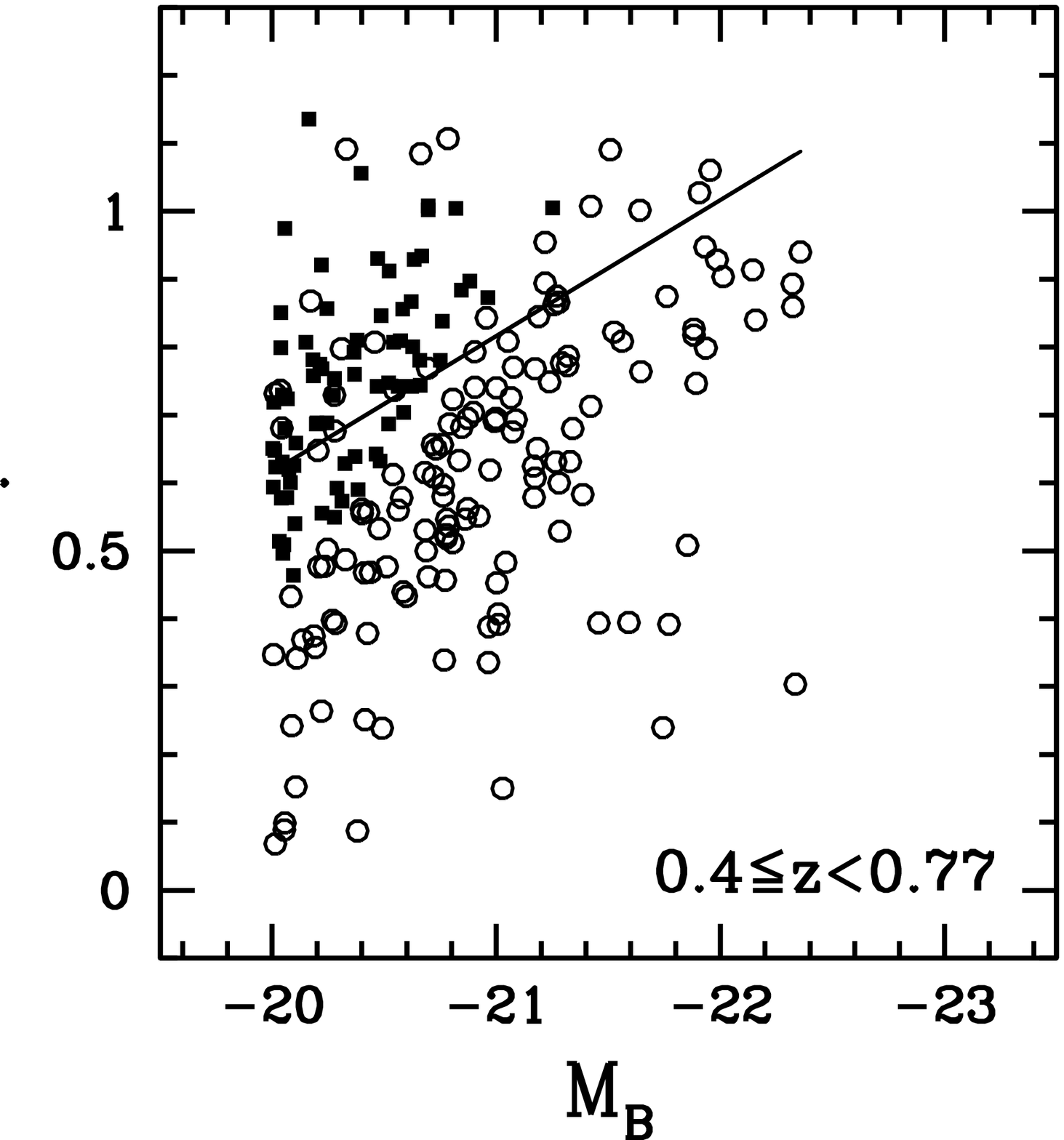}}
\mbox{\includegraphics[width=50mm]{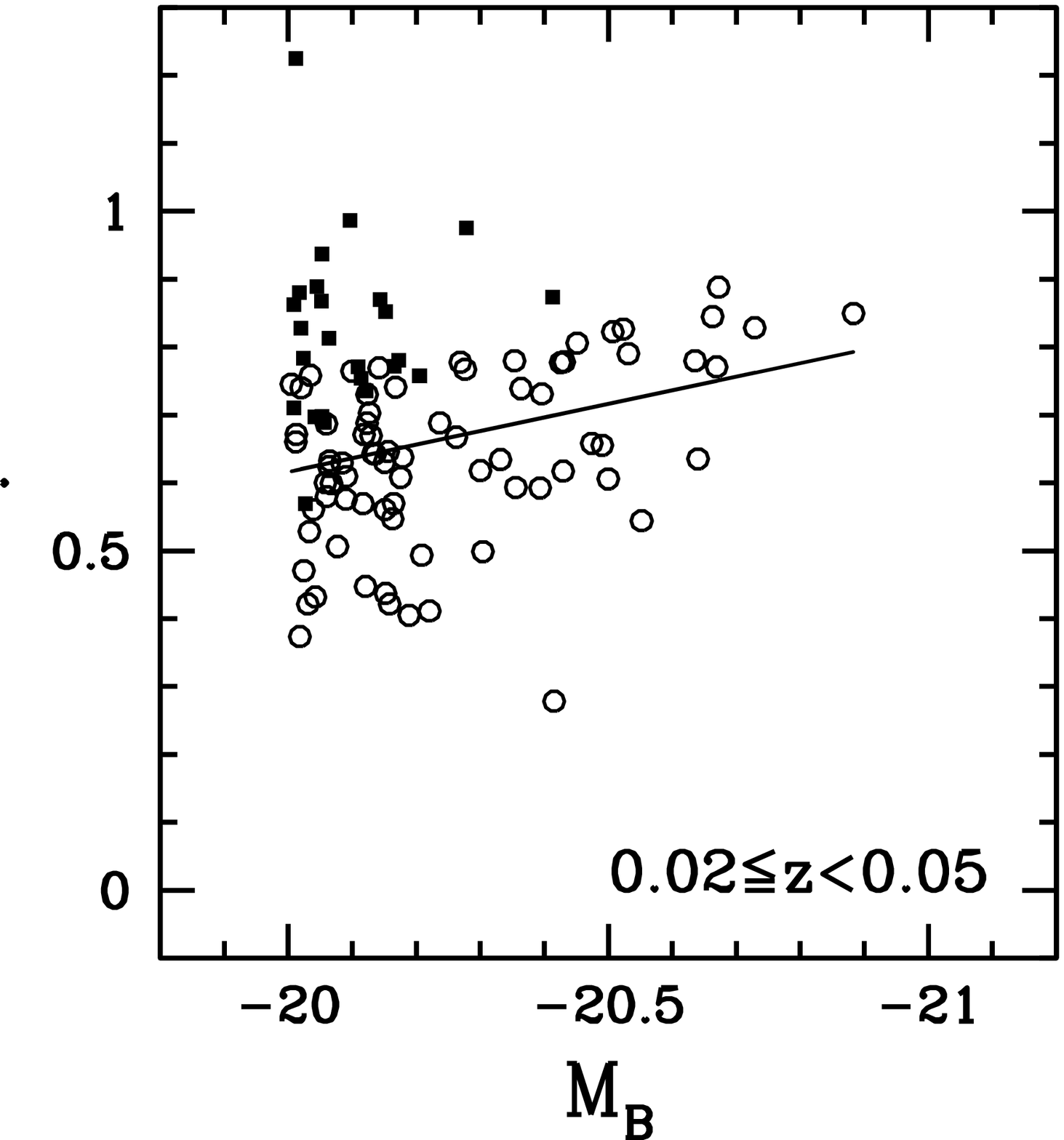}}
  \caption{The magnitude-size distribution of bulgeless (solid-squares) and normal-disc (open-circles) galaxies is shown in the three redshift ranges for rest-frame {\it B}-band. The solid line is the Freeman's relation \citep{b17}, drawn using the constant value of surface brightness of 21.65 mag arcsec$^{-2}$. The bulgeless sources populate the upper side of the solid-line i.e. they support larger sizes and hence lower brightness. For the normal-discs, the scatter is more towards smaller sizes (i.e. towards more brightness).}
\end{figure*}

\begin{figure*}
\mbox{\includegraphics[width=53.5mm]{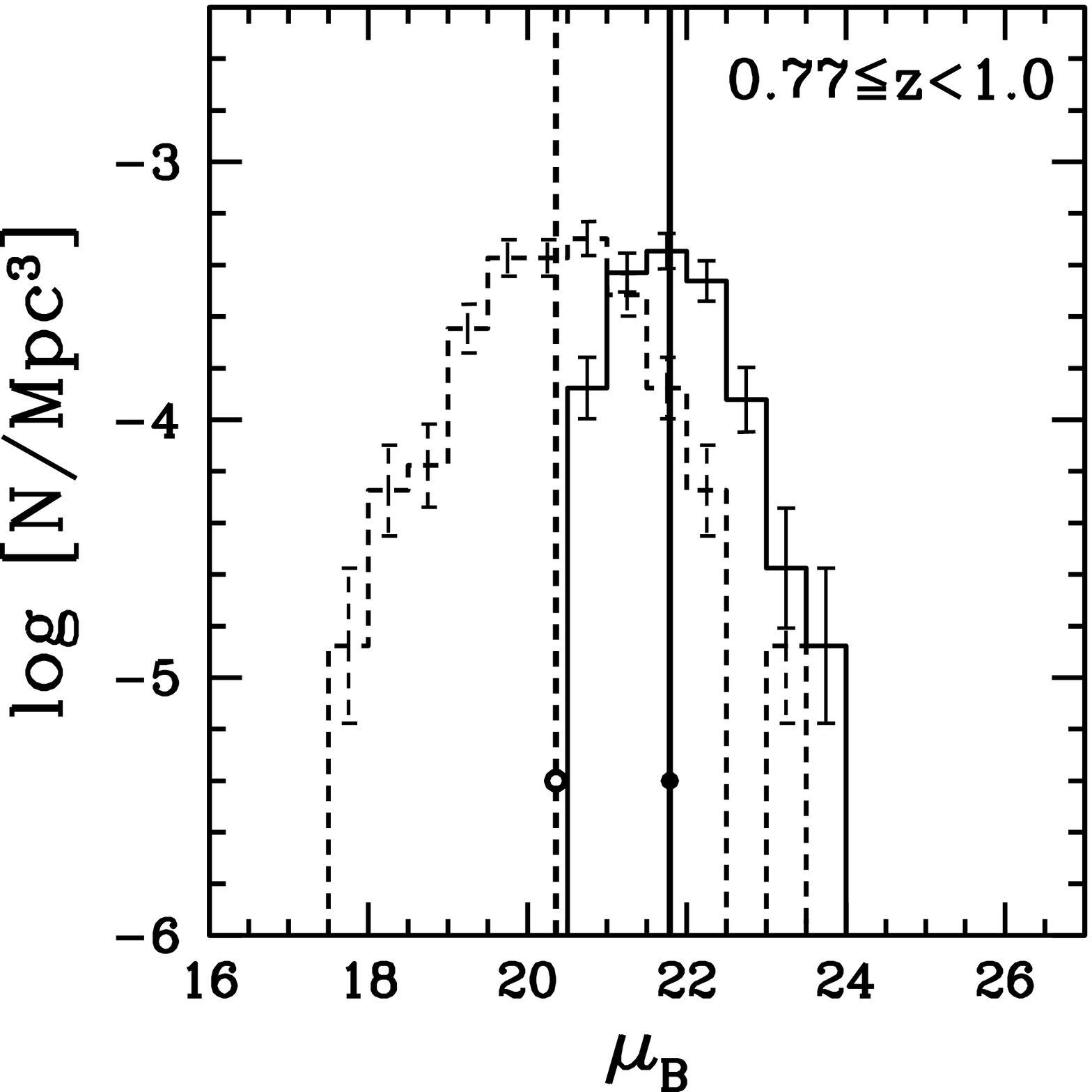}}
\mbox{\includegraphics[width=50mm]{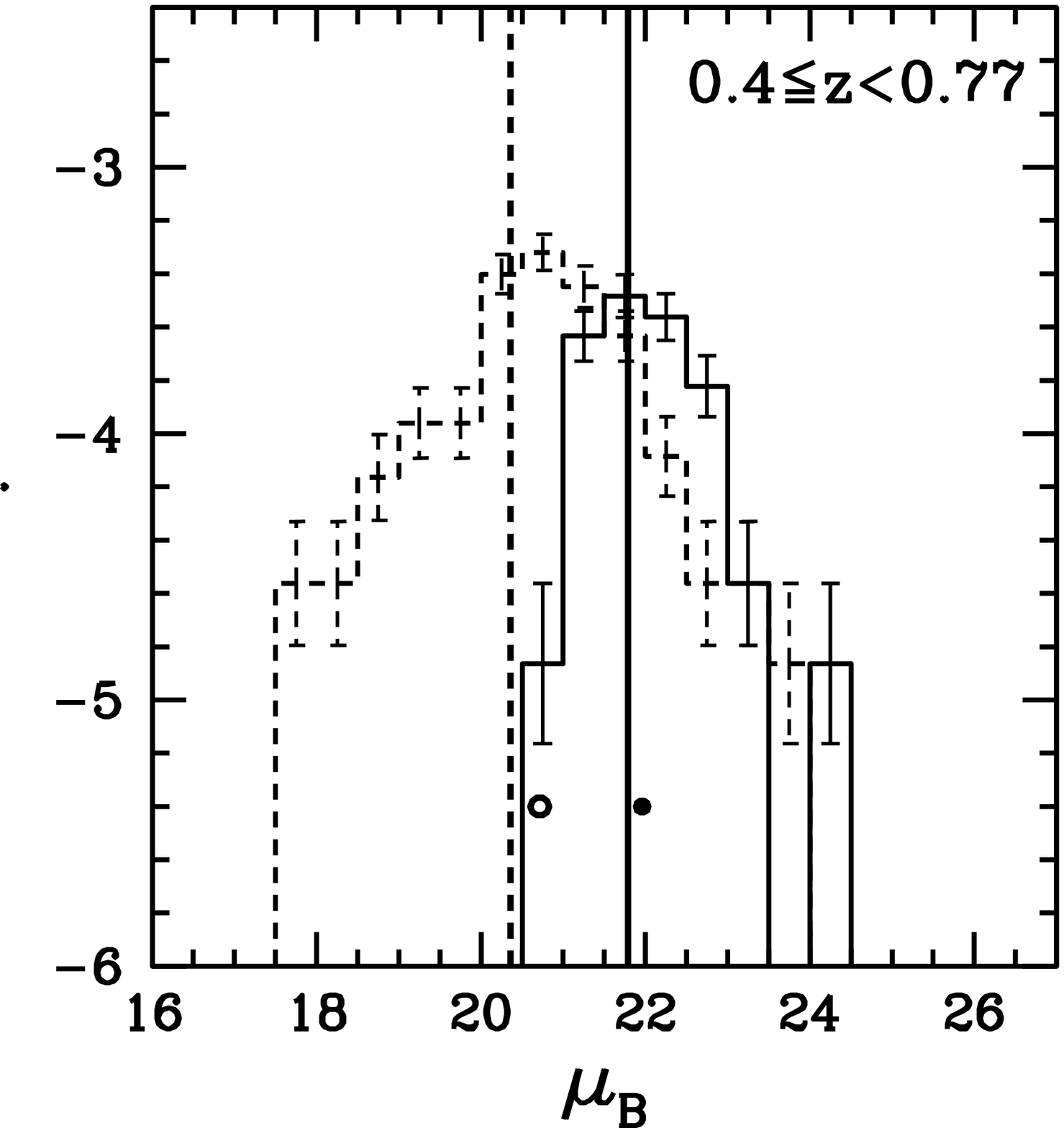}}
\mbox{\includegraphics[width=50mm]{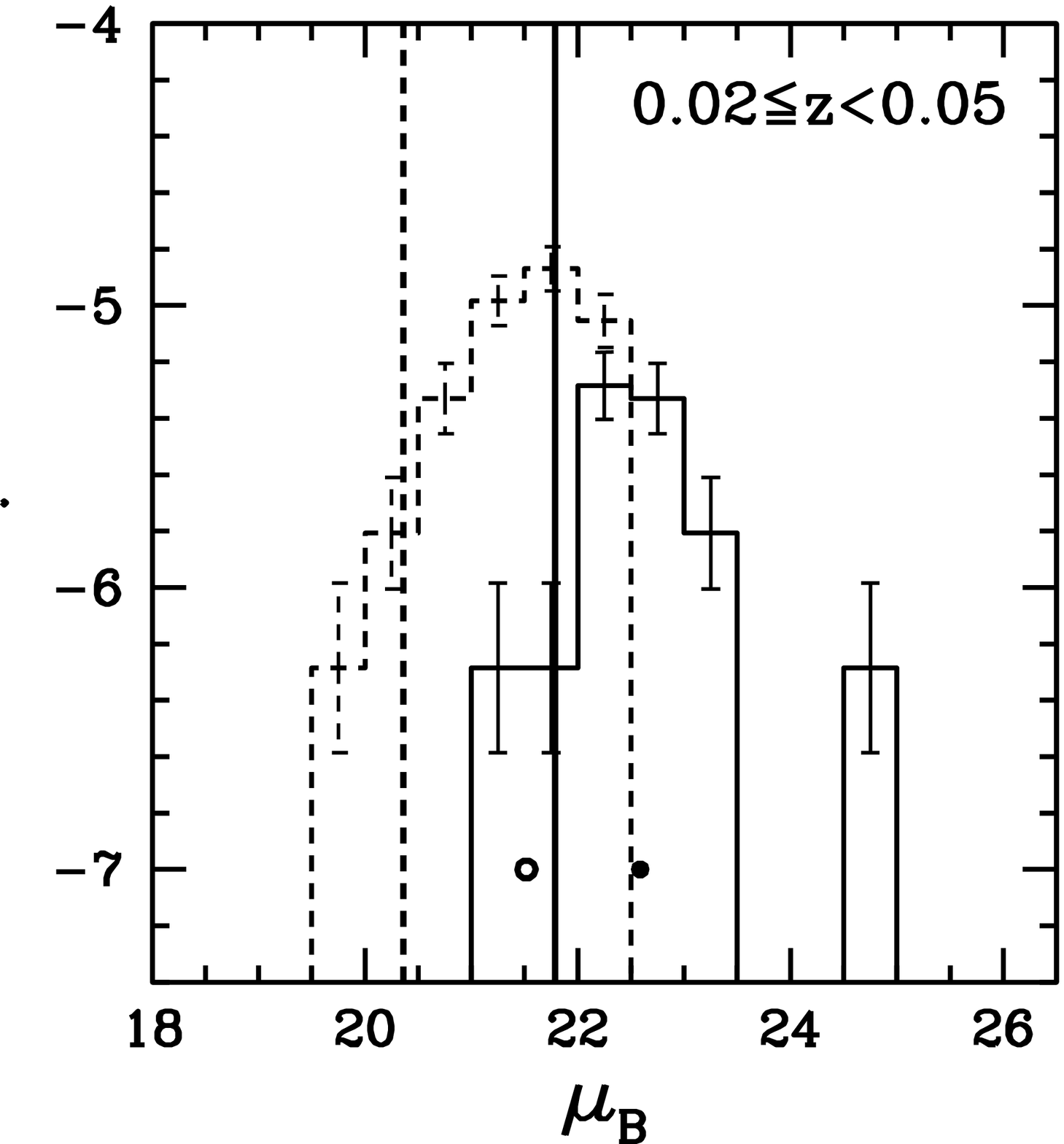}}
  \caption{The surface-brightness distribution of bulgeless (solid-lines) and normal-disc (dashed-lines) galaxies is shown in the three redshift ranges for rest-frame {\it B}-band. The solid-dot indicates the mean value of the bulgeless sources, in each redshift range, while the open-circle indicates the mean value of normal-discs. The lines passing through the means at higher z-range (0.77-1.0) are drawn as it is in the other two redshift ranges to observe evolution. Both the means have shifted towards higher $\overline{\mu}_B$ (or lower surface-brightness). The shift increases as we move from the lower z-range (0.4-0.77) to the local z-range(0.02-0.05).}
\end{figure*}

\section{Luminosity-function and size-distribution}

\subsection{Luminosity-function}

Luminosity-function, $\phi(L)$, for a set of galaxies gives the number density of such galaxies per luminosity interval. It has been used in almost every large redshift survey to determine the evolutionary relation for galaxies. \citet{b50} and \citet{b23} have explored the relative merits of different methods that are used to employ this function to understand the data.

The Schechter's form of luminosity-function \citep{b43} is,
\begin{equation}
\phi(L)dL=\phi^*(L)(L/L^*)^{\alpha}\exp(-L/L^*)dL,
\end{equation}
where $L^*$ is the characteristic luminosity at which the function exhibits a rapid change from power-law to exponential-law. $\alpha$ is the slope for the power-law and $\phi^*$ is the normalization. The form parametrized in terms of absolute-magnitude is shown:
\begin{equation}
\begin{split}
log_{10}\phi(M)& =log_{10}\phi^*+log_{10}(0.4*2.303)\\
              & \quad +0.4(\alpha+1)(M^*-M)\\
              & \quad -10^{(0.4(M^*-M))}/2.303,
\end{split}
\end{equation} 
where $\phi(M)$ is the number of galaxies per Mpc$^3$ having absolute-magnitude M. M$^*$ is the characteristic magnitude and $log_{10}(\phi^*)$ is the normalization.

The normalization, $log_{10}(\phi^*)$, has been calculated and fixed for both bulgeless and normal-discs for the three z-ranges. To compute the value, the log of number of sources per unit comoving-volume is found in each magnitude bin and mean is calculated.
 
The power-law slope gives random values if the number of sources in the higher magnitude bin is less. \citet{b51} found that $\alpha$$=$-1.3 adequately describes the sources in rest-frame {\it B}-band and provides a good description of the data even in highest redshift bins (near z$=$1). Thus, the fixed value of $\alpha$$=$-1.3 is used in the analysis. 

In Fig.~6, the density-distribution of bulgeless and normal-discs is shown in the three redshift ranges (0.77-1.0, 0.4-0.77, 0.02-0.05). Errors are computed assuming Poisson-distribution at each value. Using the two fixed parameters, $\alpha$ and $log_{10}(\phi^*)$, the characteristic magnitude, $M_B^*$, is obtained from the Levenberg-Marquardt algorithm. The algorithm provides a numerical solution to the problem of minimizing a function, over a space of parameters of the function. The function fitted for both morphological types in the three redshift-ranges is shown with solid-lines for the bulgeless and dashed-lines for the normal-discs.

The parameter values obtained in the two z-ranges and the local z-range are given in Table~3. A slight decrease is seen in the value of $M_B^*$, from higher to lower z-range, though an increase is expected. The decrease of this kind can also be seen in the values found by \citet{b51} from z$\sim$0.9 to z$\sim$0.7. From $z_{mean}$$=$0.89 to $z_{mean}$$=$0.04, the change in characteristic magnitude, $\Delta$$M_B^*$, is 0.55-mag for bulgeless sources and 0.95-mag for normal-discs. 

Thus, decrease in luminosity with time is $\sim$0.4-mag more for normal-discs as compared to the bulgeless sources. This adds to the observation that the dimming in surface-brightness is also found to be 0.4-mag more for the normal-discs (in Section~3). 

\subsection{Size distribution} 

The size evolution of galaxies with redshift serves as an important constraint on the models of galaxy evolution. The size-distribution of bulgeless and normal-discs is shown for the three redshift-ranges (Fig.~7). The mean, $\overline{\log_{10}(R_{eB})}$, and standard-deviation (std.dev.), $\sigma$($\log_{10}(R_{eB})$), values of the logarithm of sizes for bulgeless and normal-disc galaxies is given in Table~4. These parameters (mean, std.dev. and total-number) are employed to fit a log-normal curve on the distribution of sizes. 

The curves provide a good fit which shows that both bulgeless and normal-disc galaxies show log-normal distribution of sizes at all redshift ranges. This is in concordance with the hierarchical growth model expectations.  Lines passing through the mean values of bulgeless and normal-discs sources in the higher z-range plot, are drawn as it is for the lower z-range plot (Fig.~7). The solid-dot indicates the mean value of bulgeless sources, in each range, while the mean value of discs is indicated using open-circle. A shift towards larger $\overline{\log_{10}(R_{eB})}$ is apparent for both the morphological types. 

The increase in the mean value, $\Delta$$\overline{\log_{10}(R_{eB})}$, can be seen from Table~4. For the bulgeless galaxies, there is an overall increase of 0.114 from $z_{mean}$$=$0.89 to $z_{mean}$$=$0.04, in rest-frame {\it B}-band. For the normal-discs, the increase is of 0.117, almost same as that seen for the bulgeless ones. Thus, the evolution of sizes with redshift is quite similar for bulgeless and normal-disc galaxies. 

\begin{table}
 \centering
 \begin{minipage}{100mm}
  \caption{The luminosity-function parameters for {\it B}-band}
  \begin{tabular}{@{}llllllllll@{}}
  \hline
   Redshift & Disc & $log_{10}\phi^*$ & $\alpha$ & $M_B^*$\\
   range & type & (computed) & (fixed) & (computed)\\
 \hline
 0.77-1.0 & bulgeless & -3.23 & -1.3 & -20.99$\pm$0.06\\
 0.4-0.77 & bulgeless & -3.43 & -1.3 & -21.02$\pm$0.23\\
 0.02-0.05 & bulgeless & -4.27 & -1.3 & -20.44$\pm$0.09\\
 0.77-1.0 & normal & -3.36 & -1.3 & -22.07$\pm$0.19\\
 0.4-0.77 & normal & -3.59 & -1.3 & -22.38$\pm$0.21\\
 0.02-0.05 & normal & -4.65 & -1.3 & -21.12$\pm$0.02\\

 \hline
\end{tabular}
\end{minipage}
\end{table}

\begin{table}
 \centering
 \begin{minipage}{100mm}
  \caption{The size-distribution parameters for {\it B}-band}
  \begin{tabular}{@{}llllllllll@{}}
  \hline
   Redshift & Disc & Mean & Std.Dev. & in-Kpc\\
   range & type & $\overline{\log_{10}(R_{eB})}$ & $\sigma$ & $\overline{R_{eB}}$\\
 \hline
 0.77-0.4 & bulgeless & 0.709 & 0.141 & 5.117\\
 0.4-0.77 & bulgeless & 0.748 & 0.144 & 5.598\\
 0.02-0.05 & bulgeless & 0.823 & 0.125 & 6.653\\
 0.77-1.0 & normal & 0.523 & 0.218 & 3.334\\
 0.4-0.77 & normal & 0.615 & 0.234 & 4.121\\
 0.02-0.05 & normal & 0.640 & 0.127 & 4.365\\
 0.77-0.4 & full-disc & 0.597 & 0.212 & 3.96\\
 0.4-0.77 & full-disc & 0.661 & 0.217 & 4.58\\
 0.02-0.05 & full-disc & 0.686 & 0.149 & 4.85\\
 \hline
\end{tabular}
\end{minipage}
\end{table}

\begin{figure*}
\mbox{\includegraphics[width=53.5mm]{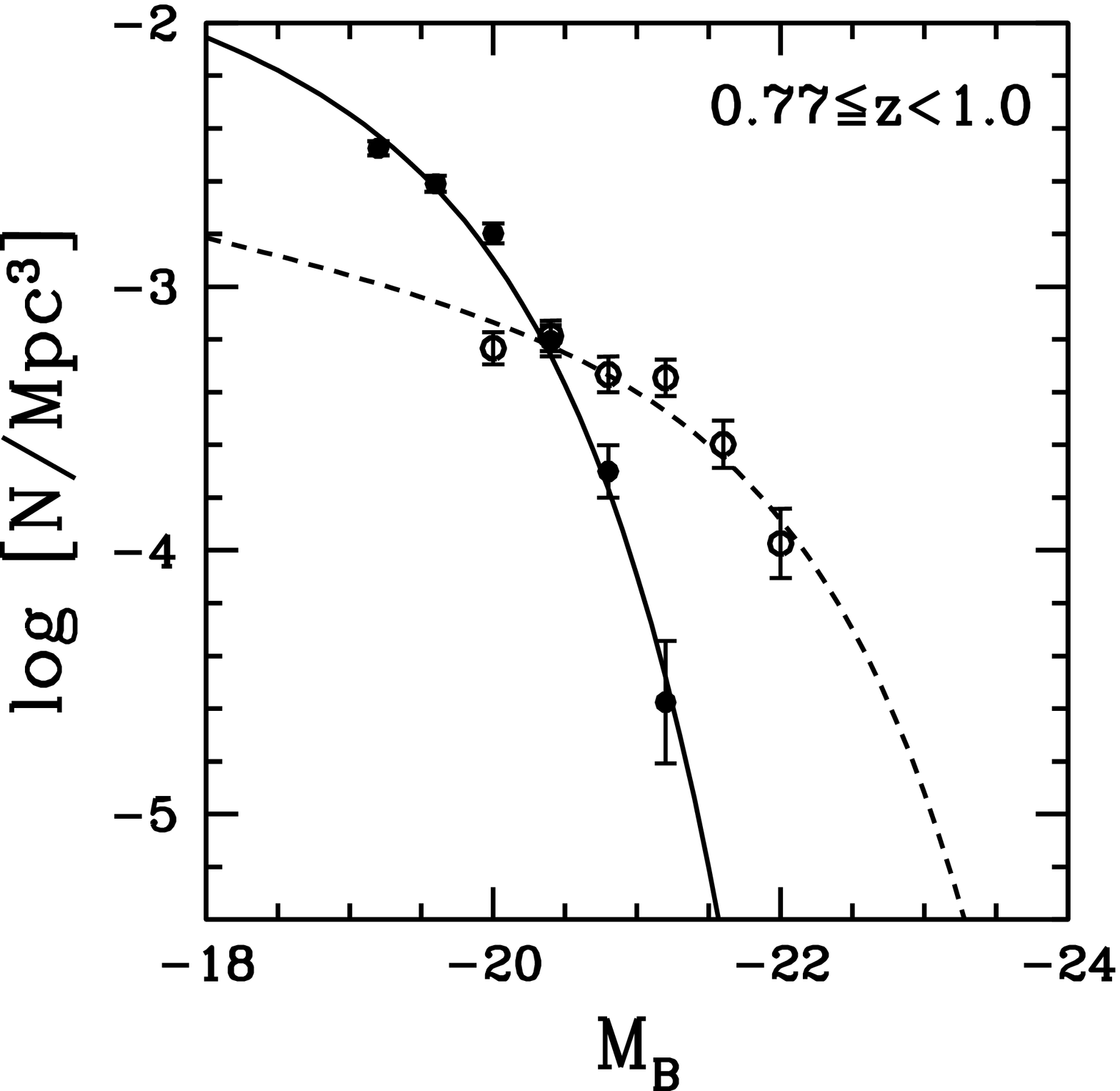}}
\mbox{\includegraphics[width=50mm]{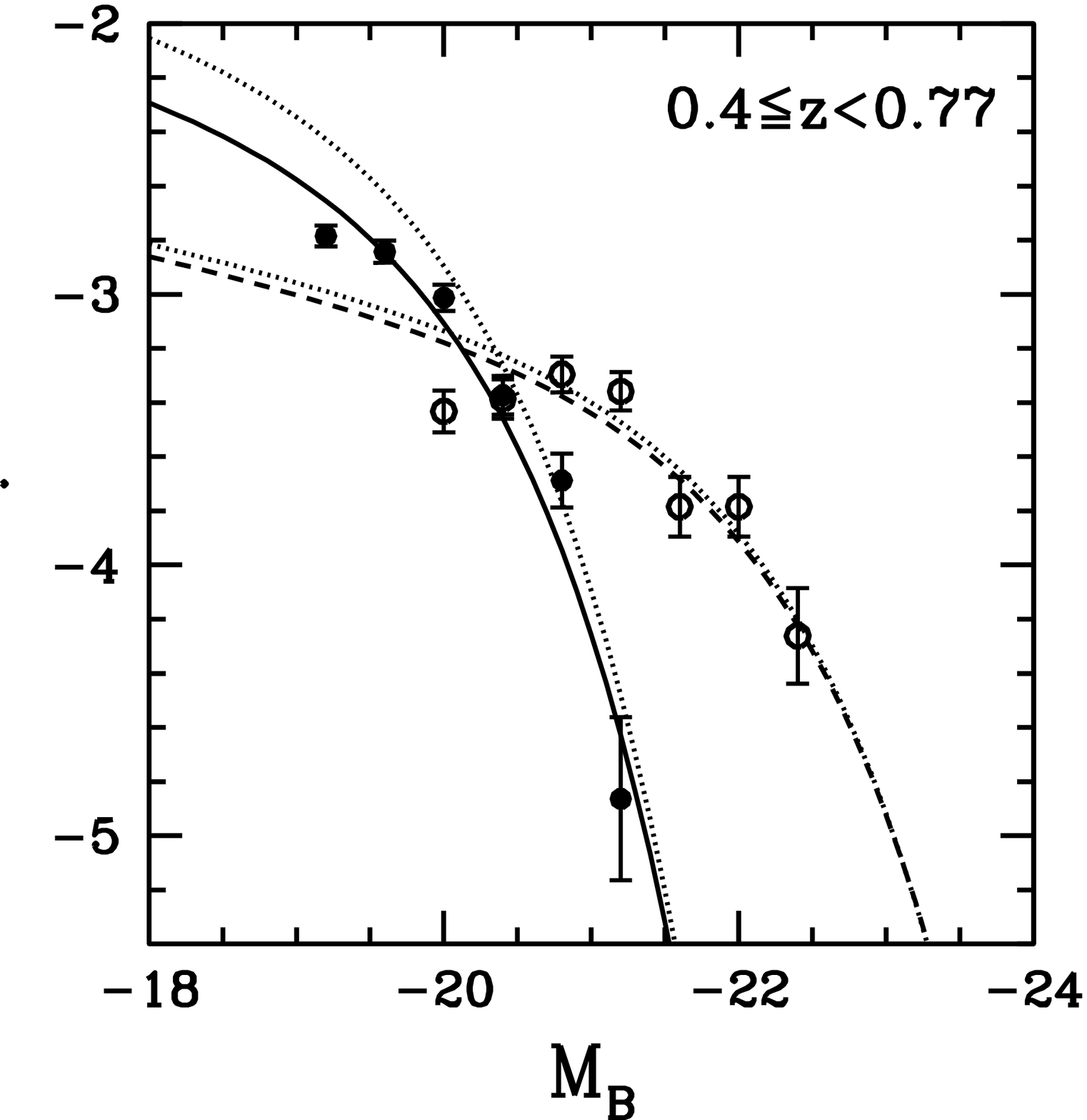}}
\mbox{\includegraphics[width=48mm]{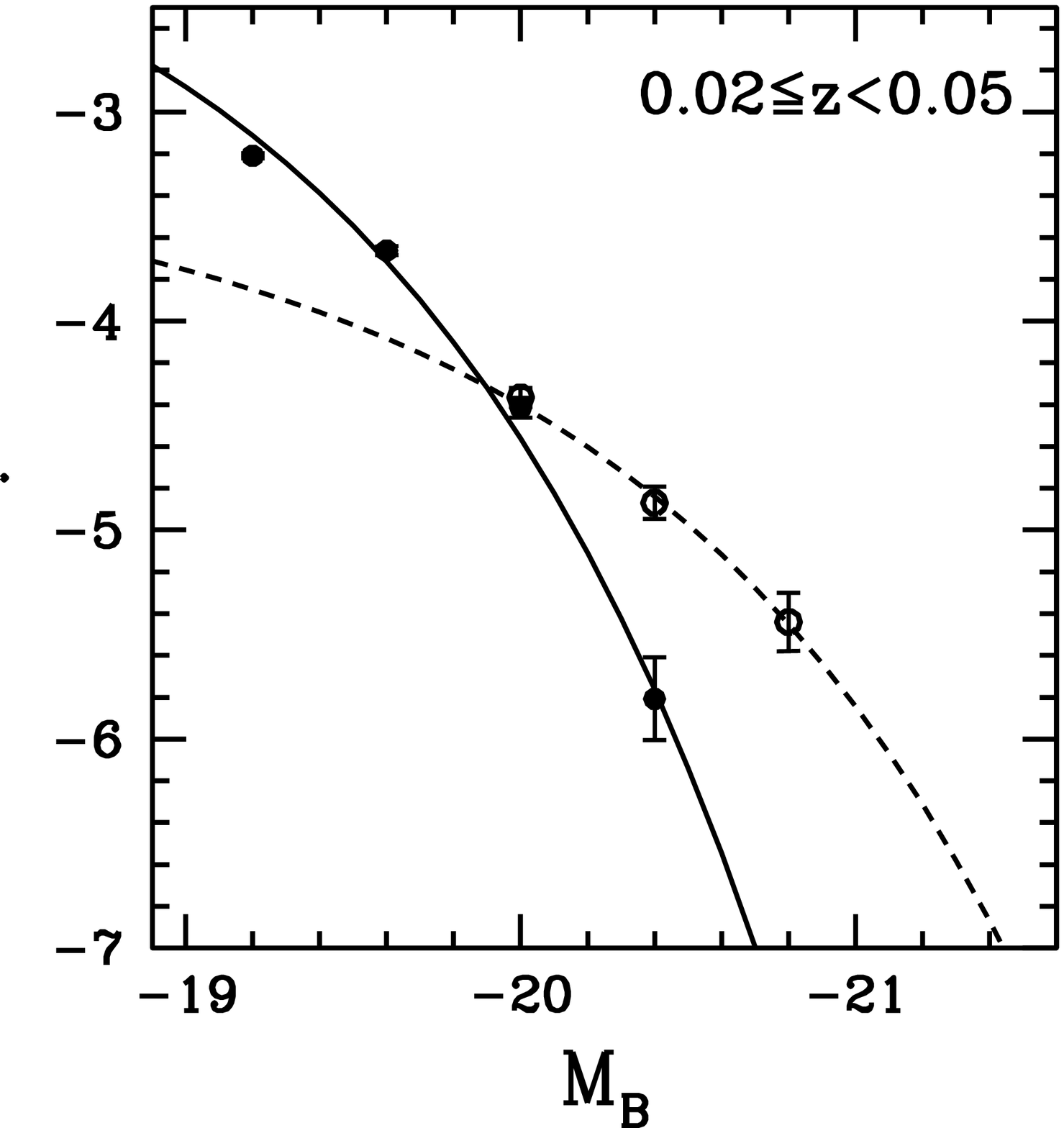}} 
  \caption{The density-distribution of bulgeless (solid-circles) and normal-disc (open-circles) galaxies is shown in the three redshift-ranges. Errors computed assuming Poisson-distribution, are shown as error-bars. The luminosity-function found after fixing $\alpha$, is fitted on the bulgeless (solid-line) and normal-disc (dashed-line) data. The fit found for bulgeless and normal-disc galaxies in the higher z-range (0.77-1.0) is held fixed (dotted-lines) for the lower z-range.}
\end{figure*}

\begin{figure*}
\mbox{\includegraphics[width=53.5mm]{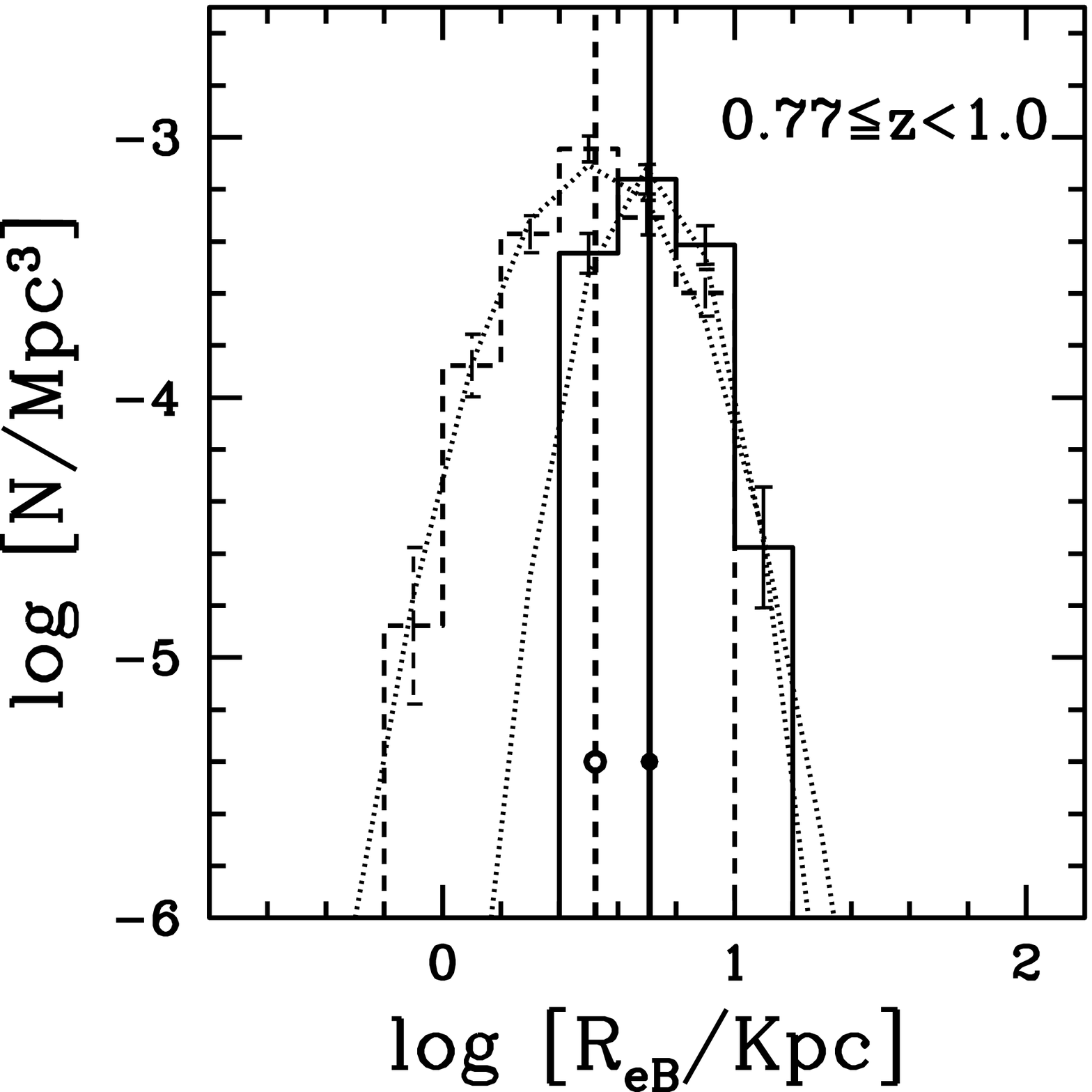}}
\mbox{\includegraphics[width=50mm]{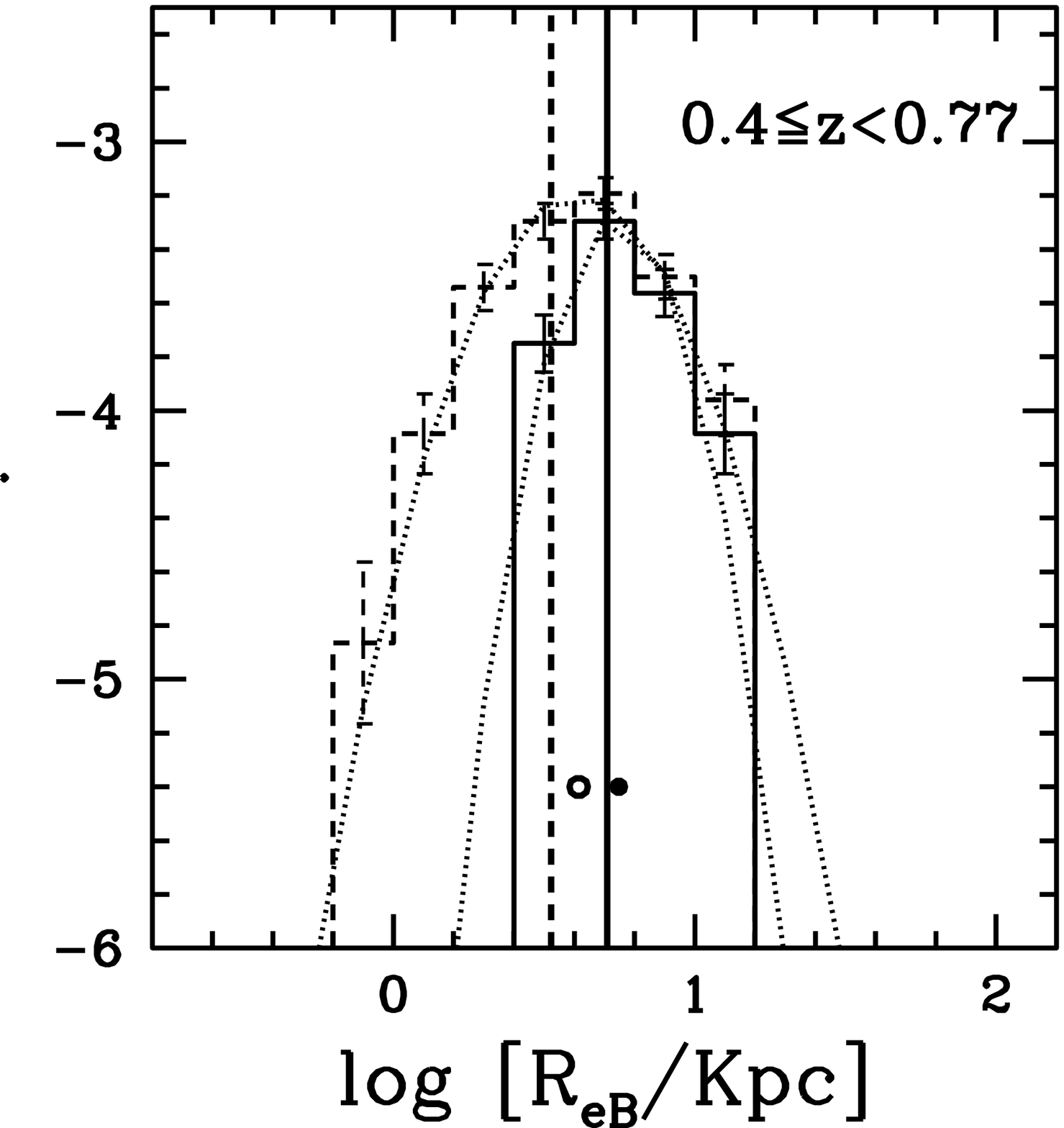}}
\mbox{\includegraphics[width=50mm]{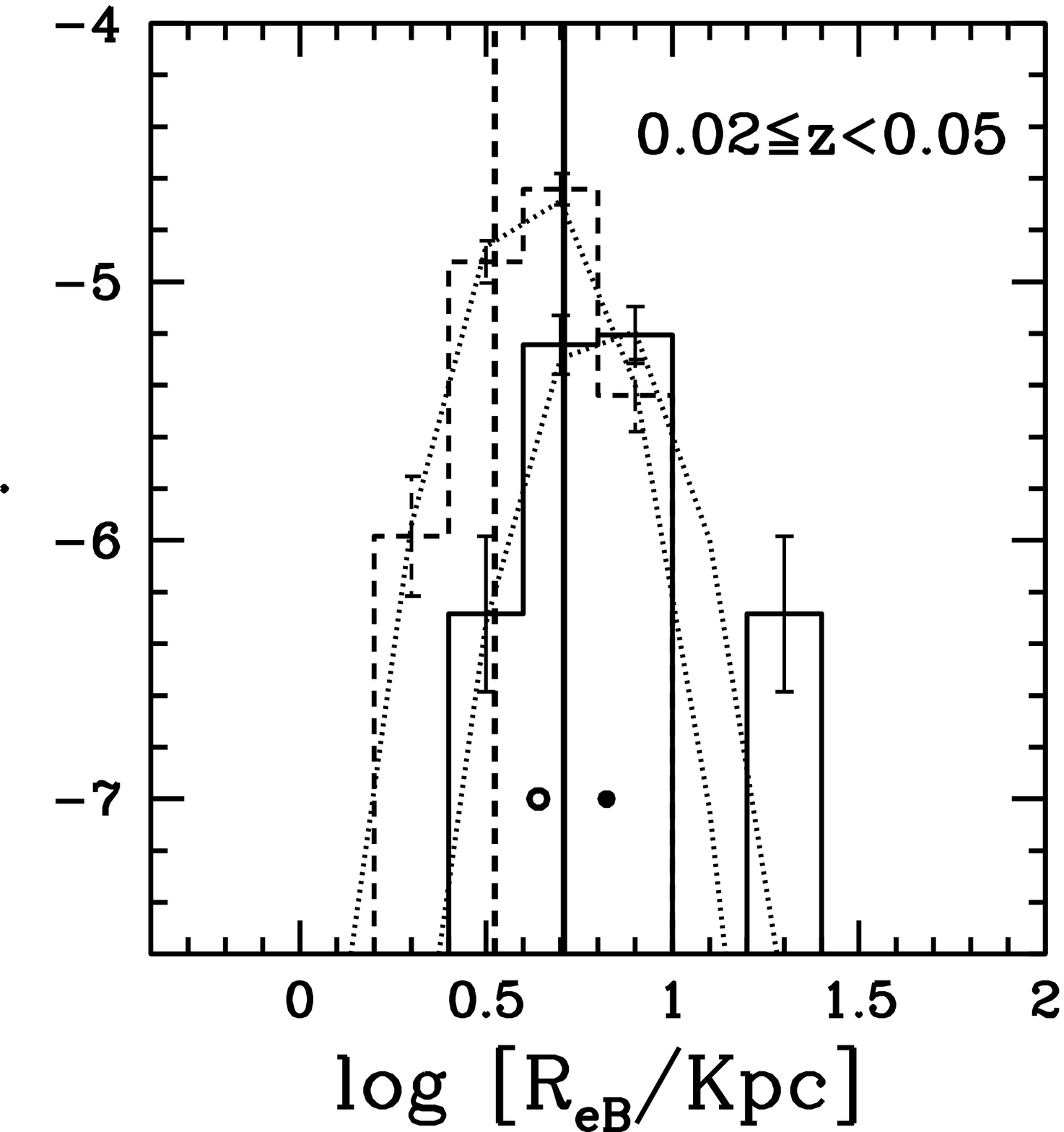}}  
  \caption{The size-distribution of bulgeless (solid-lines) and normal-disc (dashed-lines) galaxies is shown in the three redshift-ranges. Errors computed assuming Poisson-distribution, are shown as error-bars. The sizes show log-normal distribution as the curves created using mean and standard-deviation values fit well to the data (dotted-lines). The solid-dot marks the mean value of the bulgeless sources, in each redshift-range, while the open circle marks the mean value of the normal-discs. The lines passing through the means at higher z-range (0.77-1.0) are drawn as it is in the other two redshift ranges to observe evolution. Both the means show a shift towards larger sizes. The shift in the mean value seems similar for the two morphological types.}
\end{figure*}

\section{Comparison between rest-frame {\it B} and {\it I}-band}

For the 0.4-1.0 redshift-range we obtained 4124 galaxies in the rest-frame {\it B}-band. There were only 727 sources out of this sample which were found to have $M_B$$\leq$-20. For this redshift range (0.4-1.0), a catalog of 1109 galaxies was obtained in the rest-frame {\it I}-band. The two catalogues in rest-frame {\it B} and {\it I}-band are matched using RA and Dec, with a maximum distance of 0.5 arc-seconds. We obtain a common sample of 174 galaxies (0.4$\leq$z$<$1.0, $M_B$$\leq$-20). This is used to compare the absolute-magnitude, half-light-radius and sersic-index of the galaxies in the two bands.    

All the galaxies (172 out of 174) are brighter in the {\it I}-band. The half-light-radius is larger in the {\it B}-band for $\sim$75$\%$ of the galaxies. The sersic-index is higher in the {\it I}-band for $\sim$73$\%$ of the galaxies, as is evident from Fig.~8. The argument that the bulge component shows itself more prominently in the {\it I}-band and the disc structure is visible in full length in the {\it B}-band, supports these statistics. 

Even for those galaxies which have a lower sersic-index in the {\it I}-band, the half-light radius is larger in the {\it B}-band for most of them (88$\%$). These discs seem to be supporting a young stellar bulge. 

The sersic-indices of the galaxies in the two rest-bands is compared in Fig.~9. It is seen that the values are quite comparable. Though the sersic-index is higher in the {\it I}-band, the difference is minor enough to not affect the morphological classification. The dashed lines marking $n_B$$=$1.7 and $n_I$$=$1.7 show that for most of the galaxies with sersic-index less than 1.7 in {\it B}-band, it is less than 1.7 in the {\it I}-band as well. Thus, the stellar density is almost tracking the mass density in galaxies, which makes it quite reasonable to measure and understand morphology in the optical band.

The distribution of the color of the common set of galaxies with redshift is shown (Fig.~10). For the galaxies which are bulgeless in the {\it B}-band, 62$\%$ are found to be bluest ($M_B$-$M_I$$<$1). The percentage reduces to $\sim$50$\%$ for the galaxies with-bulge. For the small number of galaxies which are reddest ($M_B$-$M_I$$>$2), total 9 in number, only 2 of them are bulgeless. Thus, the bulgeless population is slightly bluer than the population with bulge.  

Fig.~11 and Fig.~12 show the images of some of the galaxies from the common set. These galaxies are bulgeless in the {\it B}-band. The upper row shows the images in rest-frame {\it B}-band (observed in {\it z}-band), and the lower row shows the same sources in the rest-frame {\it I}-band (observed in {\it H}-band). The galaxies belong to the 0.8-1.0 redshift-range. The three galaxies shown in Fig.~11 have $R_{eB}$$<$$R_{eI}$ while those shown in Fig.~12 have $R_{eB}$$>$$R_{eI}$. The bulge seems more prominent and the structure seems smoother in the {\it I}-band.

\begin{figure}
\mbox{\includegraphics[width=70mm]{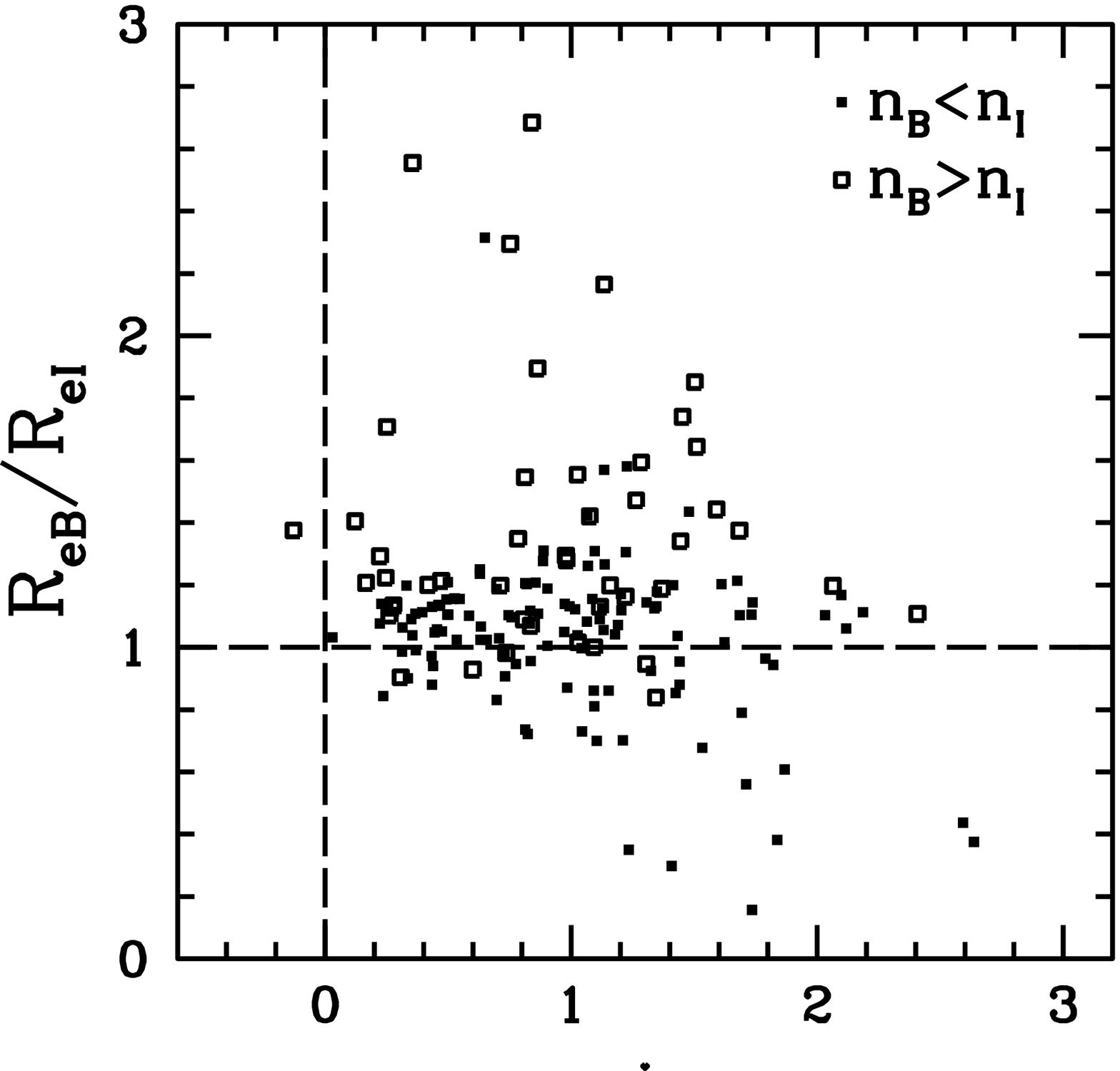}}
\mbox{\includegraphics[width=70mm]{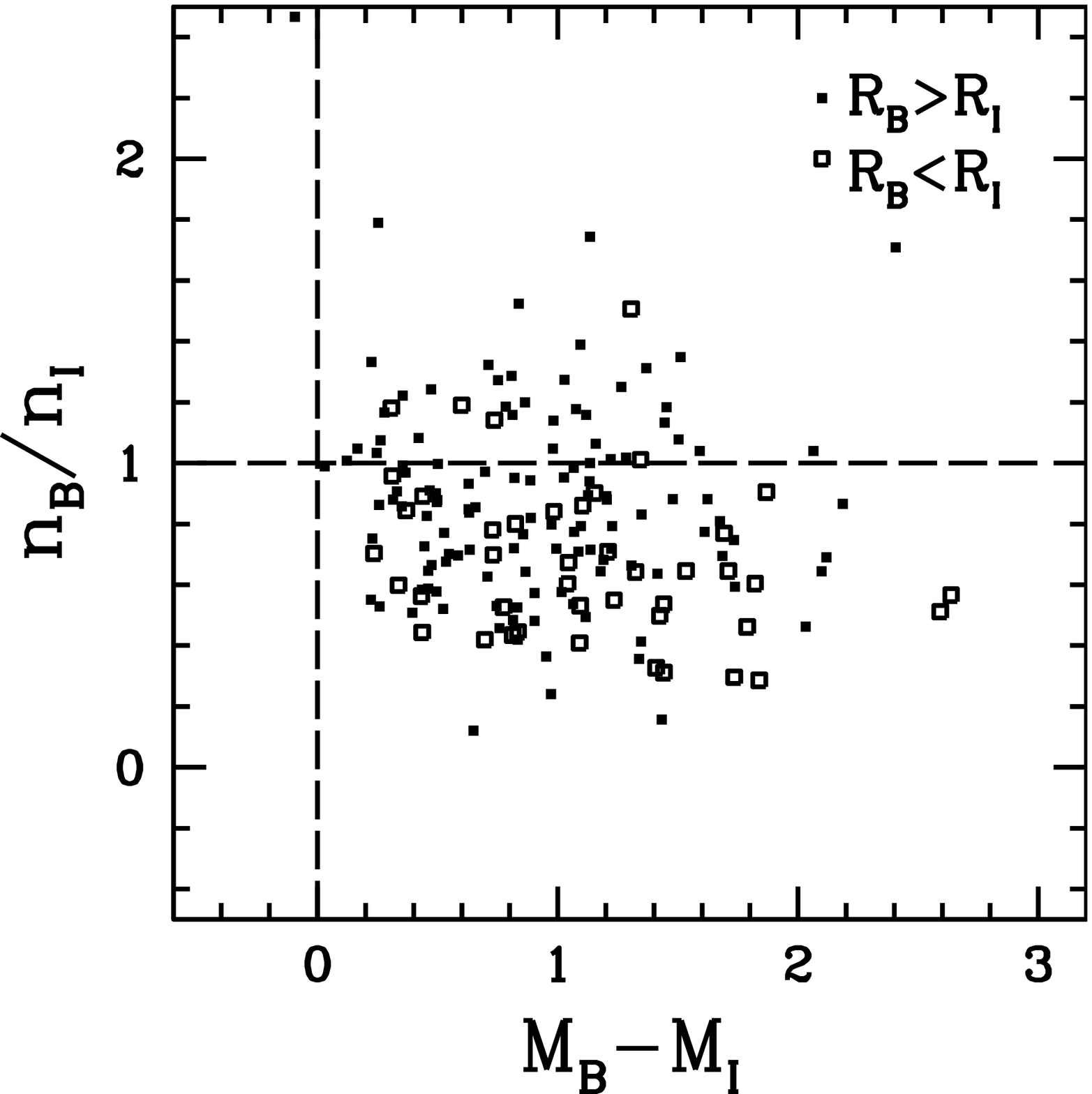}}
  \caption{The ratio of sizes, difference of absolute-magnitudes and ratio of sersic-indices of the galaxies in {\it B} and {\it I}-band are used to find the correlation among the values in the two bands. All the galaxies (172 out of 174) have ($M_B$-$M_I$)$>$0. Most of the galaxies ($\sim$75$\%$) have ($R_{eB}$/$R_{eI}$)$>$1 as seen in the first plot. The second-plot shows that ($n_B$/$n_I$)$<$1 for most ($\sim$73$\%$) of the galaxies.}
\end{figure}

\begin{figure}
\mbox{\includegraphics[width=70mm]{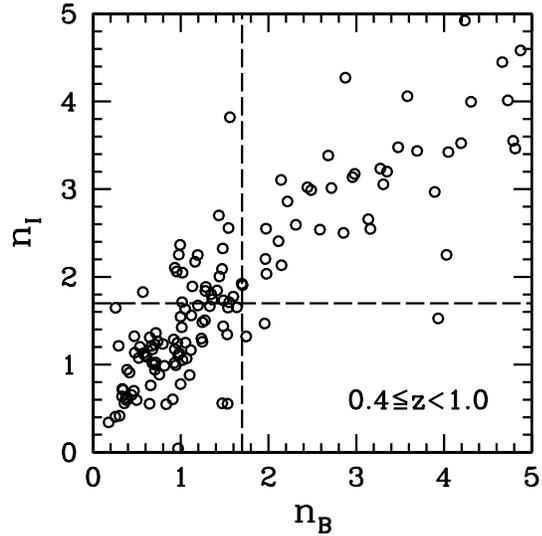}}
  \caption{The distribution of sersic-indices of the galaxies in {\it B} and {\it I}-band is shown for the common sample. For the galaxies with $n_B$$<$1.7, most of them ($\sim$73$\%$) have $n_I$$<$1.7 also.}
\end{figure}

\begin{figure}
\mbox{\includegraphics[width=70mm]{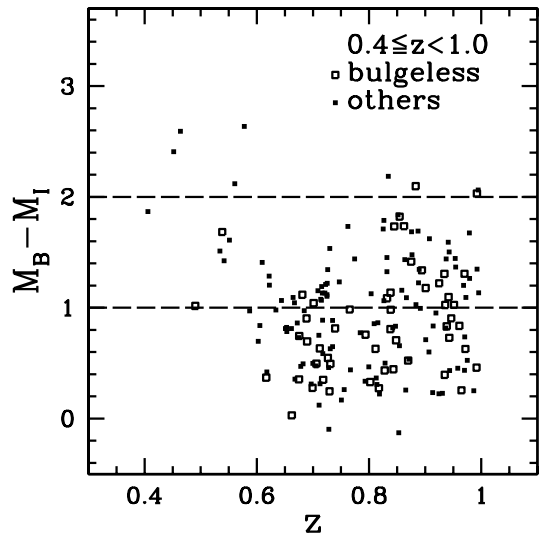}}
  \caption{The distribution of the magnitude-difference of galaxies with redshift is shown for the common sample. More than 62$\%$ of the bulgeless galaxies are found to be bluest ($M_B$-$M_I$$<$1). Reddest ($M_B$-$M_I$$>$2) of the galaxies are with-bulge and observed at lower redshifts (0.4-0.6).}
\end{figure}

\begin{figure}
\mbox{\includegraphics[width=84mm]{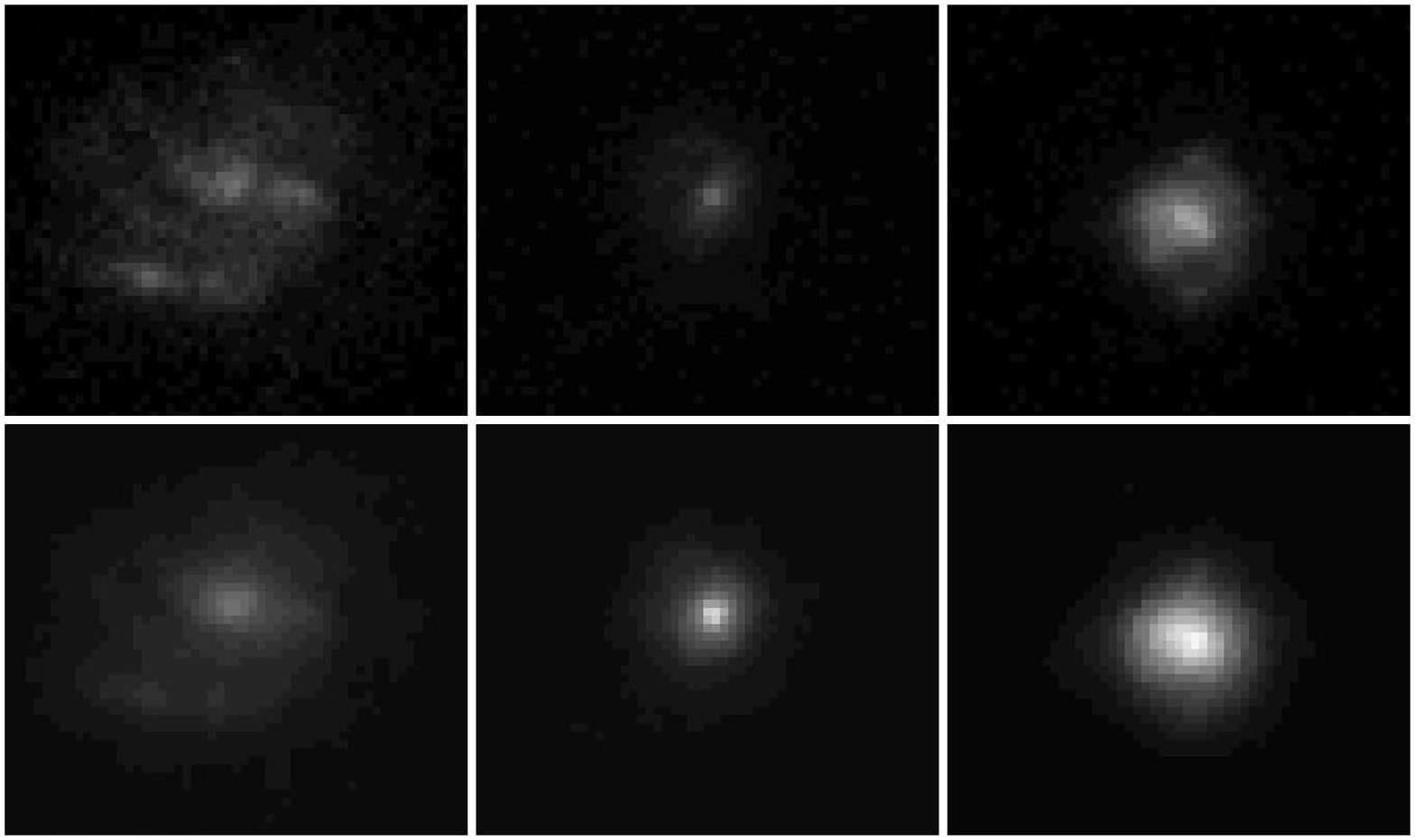}}
  \caption{The images of galaxies from the common-set that are bulgeless in the {\it B}-band. The upper row shows the images in rest-frame {\it B}-band, and the lower row shows the same galaxies in rest-frame {\it I}-band. The galaxies are face-on and have $R_{eB}$$<$$R_{eI}$. The images in {\it I}-band show them with a smoother structure and a more prominent bulge.}
\end{figure}

\begin{figure}
\mbox{\includegraphics[width=84mm]{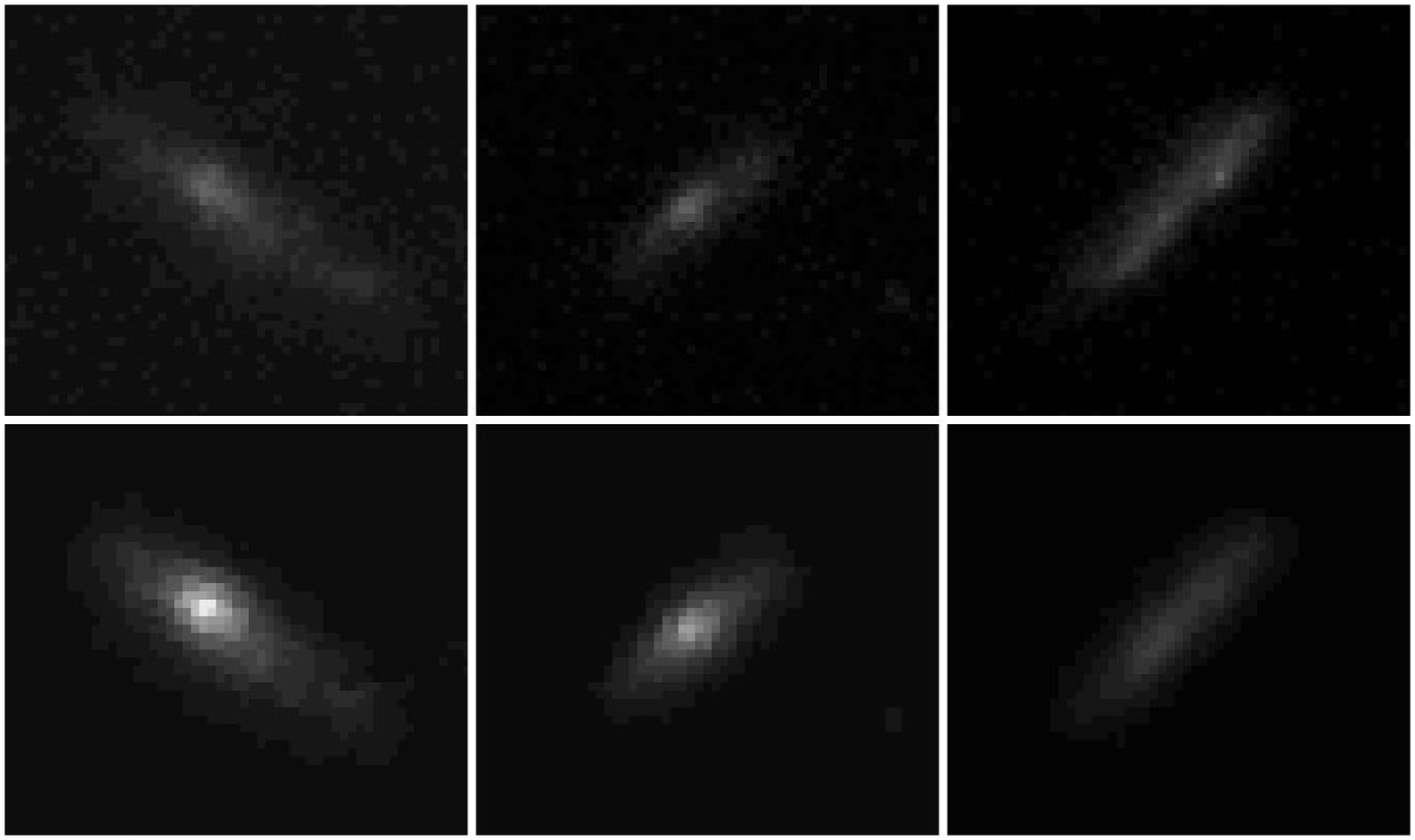}}
  \caption{The images of galaxies from the common-set that are bulgeless in the {\it B}-band. The upper row shows the images in rest-frame {\it B}-band, and the lower row shows the same galaxies in rest-frame {\it I}-band. The galaxies are edge-on and have $R_{eB}$$>$$R_{eI}$. The images in {\it I}-band show them with a smoother structure and a more prominent bulge.}
\end{figure}

\section{Discussion and conclusions}

The presence of bulgeless discs in the universe pose a problem for the structure formation scenario given by $\Lambda$CDM cosmology in which mergers and accretions play a major role. Examining the surface-brightness, luminosity and size evolution of such bulgeless galaxies vis-a-vis normal-disc galaxies can help understand the reasons for their presence and survival with time. We have studied the evolution of bright ($M_B$$\leq$-20) bulgeless and normal-disc galaxies in the rest-frame {\it B}-band in three redshift ranges (0.02$\leq$z$<$0.05, 0.4$\leq$z$<$0.77, 0.77$\leq$z$<$1.0).

In the magnitude-size distribution plots, it is seen that the sizes are larger for the bulgeless galaxies as compared to the normal-discs. Though the bulgeless sources support larger sizes, the population is confined to low luminosities. With decreasing redshifts, this pattern becomes more pronounced. Overall it suggests that the average surface-brightness is lower for the bulgeless galaxies. In the surface-brightness distribution, it is seen that this is indeed the case. The discs with bulge are brighter than the bulgeless ones by more than 1-mag at all redshift-ranges.  

In the surface-brightness evolution, it is seen that the dimming observed for normal-discs is $\sim$0.4-mag more than that seen for the bulgeless galaxies. While for the bulgeless galaxies the dimming, from $z_{mean}$$=$0.89 to $z_{mean}$$=$0.04, is of 0.79 mag arcsec$^{-2}$, for normal-discs it is of 1.16 mag arcsec$^{-2}$. For the disc sample as a whole, the overall dimming over the same range is of 0.86 mag arcsec$^{-2}$. Thus, the contribution to the dimming seen for the entire disc sample is more from the normal-discs than from the bulgeless sources. 

\citet{b46} applied a magnitude cut of $M_B$$<$-19 and found that there is no significant evolution in the surface-brightness of disc galaxies. \citet{b37} confirmed that no discernible evolution remains in the surface-brightness of disc-dominated galaxies once selection effects are removed. However, \citet{b2} found a brightening of $\sim$1 mag arcsec$^{-2}$ in disc galaxies by z$\sim$1 for $M_V$$\leq$-20. They emphasized that the lack of evolution in earlier studies is due to the hard lower surface-brightness cut that leads to the removal of substantial numbers of galaxies in the low-redshift bins. We have applied the magnitude cut $M_B$$\leq$-20, according to the redshift and imaging limits. The results are in agreement with significant evolution, as for the whole disc sample, dimming of $\sim$0.9-mag is seen up to z$\sim$0.9. It is an added fact that the dimming is largely from normal discs i.e. discs with bulge.

The proportion of bulgeless galaxies in the full disc sample declines with decreasing redshift. For high-z bin (0.77-1.0), 39.7$\%$ of the discs are bulgeless, for low-z bin (0.4-0.77), it reduces to 34.7$\%$ and for local-z bin (0.02-0.05), it further reduces to 24.8$\%$. The change in morphology for some fraction of the bulgeless galaxies as we approach the present epoch may explain the considerable decline in their space-density and the decrease in their proportion with respect to the full disc sample.

It can also be the probable reason for the larger amount of dimming seen in normal-discs as compared to the bulgeless galaxies. If the bulgeless ones, which are less bright than the normal-discs, switch over to the normal-disc sample with time, the normal-disc sample is expected to show a larger amount of dimming. To be able to substantiate the argument, the luminosity and size evolution of the two morphological types needs to be understood. 

To examine luminosity evolution, the Schechter's function form is fitted on the density-distribution of bulgeless and normal-disc galaxies in the three redshift-ranges. The increase in characteristic-magnitude, $M_B^*$, for bulgeless galaxies from $z_{mean}$$=$0.89 to $z_{mean}$$=$0.04, is of 0.55-mag. For the normal-discs, the increase is of 0.95-mag, that is 0.4-mag more than that seen for bulgeless galaxies. The difference in the luminosity evolution of bulgeless and normal-discs is similar to the difference in their surface-brightness evolution. The faintness and dimming, with time, observed in the case of normal-discs is $\sim$0.4-mag more than the faintness and dimming observed in the case of bulgeless galaxies. 

In the size evolution, it is seen that the sizes show a log-normal distribution for both bulgeless and normal-disc galaxies in the three redshift-ranges. This is in concordance with the hierarchical growth model expectations and has earlier been observed for disc dominated sample as a whole \citep{b45,b37}. The increase in the mean value, $\Delta$$\overline{\log_{10}(R_{eB})}$, from $z_{mean}$$=$0.89 to $z_{mean}$$=$0.04, is of 0.114 for the bulgeless galaxies. For the normal-discs, the increase is of 0.117. The evolution of sizes with redshift is almost same for bulgeless and normal-disc galaxies. The bulgeless thus seem to be witnessing the same amount of accretion as the normal-discs.

In addition to studying the evolution, the magnitude, size and sersic-index values of the galaxies in rest-frame {\it B} and {\it I}-band (for 0.4$\leq$z$<$1.0) are compared. It is argued that the surface-brightness profile observed in rest-frame {\it B}-band may not reliably trace the surface-density profile in stellar-mass. Also, for the surface-density profile, {\it I}-band should be a better choice being freer of biases of young stellar population and dust. Thus, to explore the differences in surface-brightness and surface-density profile, comparisons with the properties in rest-frame {\it I}-band are made.

For most of the galaxies, sersic-index is slightly higher in the {\it I}-band and half-light-radius is larger in the {\it B}-band. The argument that the bulge component shows itself more prominently in the {\it I}-band and the disc structure is visible in full length in the {\it B}-band supports these findings. 

Though the sersic-index for the galaxies is higher in the {\it I}-band, the variation is minor enough to not affect the morphological classification. It reflects that the stellar density is almost tracking the mass density in galaxies. Thus, it is quite reasonable to measure and understand morphology in the optical band. The bulgeless galaxies are found to be slightly bluer ($M_B$-$M_I$$<$1) as compared to the galaxies with-bulge. This is expected because of the absence of bulges which are mostly populated with older stars. 

The major conclusions of the comparative study of bulgeless and normal-disc galaxies done in the optical over three redshift ranges have been that with decreasing redshift
\begin{enumerate}
\item the dimming and faintness observed for normal-discs is more than that for bulgeless galaxies,
\item there is a decrease in the proportion of bulgeless galaxies in the full disc sample,
\item the increase in sizes for both morphological types is found to be almost similar.
\end{enumerate}
The stated outcomes support the argument that a fraction of the bulgeless galaxies witnessed bulge formation with time and switched to the normal-disc sample. It has been found, through theoretical models, that minor mergers and accretion leads to bulge growth in discs as the stellar content of the satellite simply gets added to the bulge of the primary \citep{b47,b25,b34}. The fact that the rest of the bulgeless population remains intact even after being subjected to similar conditions, needs to be understood.

As a next step, it is intended to bring more clarity to the difference in evolution of bulgeless and normal-disc galaxies. The asymmetry and clumpiness of stellar light, along with it's concentration, helps analyze the merger and accretion history of galaxies \citep{b12}. We shall thus know if the continuous accretion has been responsible for the bulge growth in a fraction of bulgeless galaxies with time. Also, the fraction and numbers can be matched with existing results to ascertain the findings.  

\section*{Acknowledgments}

This work is supported by grant 09/045(0972)/2010-EMR-I from Human Resource Development Group (HRDG), which is a division of Council of Scientific and Industrial Research (CSIR), India. The author would like to thank Swara Ravindranath for suggesting the problem and for making available the data from GOODS {\it HST}-ACS and ERS WFC3. The author would also like to thank H.P. Singh for useful discussions. Lastly, the author is thankful to the reviewer, Jason Melbourne, whose comments have helped in the major improvement of this work.

\bsp

\label{lastpage}

\end{document}